\documentclass[letterpaper,twocolumn,10pt]{article}
\usepackage{usenix,epsfig,endnotes}
\usepackage{float}
\usepackage{subfigure}
\usepackage{longtable}
\usepackage{booktabs}
\usepackage{graphicx}
\usepackage{url}
\usepackage{amsthm}
\usepackage{xcolor}
\usepackage{amsmath,amssymb,latexsym}
\usepackage{multirow}
\usepackage{tabularx}
\usepackage{epstopdf}
\usepackage{verbatim}
\usepackage[linesnumbered,ruled,figure]{algorithm2e}
\usepackage{times}
\usepackage{multirow}
\usepackage{listings}

\usepackage{thmtools}
\usepackage{thm-restate}

\definecolor{kugray5}{RGB}{224,224,224}
\definecolor{mygreen}{RGB}{0,139,0}


\newcommand{\para}[1]{{\vspace{2pt} \bf \noindent #1 \hspace{1pt}}}

\newcommand{\tf}{{TensorFlow}}
\newcommand{\pn}{{NGra}} % pn=Paper_Name

\begin{document}

%don't want date printed
\date{}

%make title bold and 14 pt font (Latex default is non-bold, 16 pt)
\title{Towards Efficient Large-Scale Graph Neural Network Computing}
%NeuGraph: an Efficient System for Graph Neural Networks in a Deep Learning Framework

%for single author (just remove % characters)
\author{
{\rm Paper \#79}\\
%{\rm Your N.\ Here}\\
%Your Institution
%\and
%{\rm Second Name}\\
%Second Institution
% copy the following lines to add more authors
% \and
% {\rm Name}\\
%Name Institution
} % end author

\author{
{\rm Lingxiao Ma$^{\dag*}$, Zhi Yang$^{\dag}$\thanks{\ These authors contributed equally to this work. \newline \hspace*{6.4pt} ${\ }^{**}$ This work is done when Lingxiao Ma is an intern and Zhi Yang is a visiting researcher at Microsoft Research.}\hspace{1ex}, Youshan Miao$^\ddag$, Jilong Xue$^\ddag$, Ming Wu$^\ddag$, Lidong Zhou$^\ddag$, Yafei Dai$^\dag$} \\
\ \\
$^\dag$Peking University, Beijing, China \\
$^\ddag$Microsoft Research, Beijing, China \\
}
%\thanks{$*$ These authors contributed equally to this work.}

\maketitle

% Use the following at camera-ready time to suppress page numbers.
% Comment it out when you first submit the paper for review.
\thispagestyle{empty}

\subsection*{Abstract}
Recent deep learning models have moved beyond
low-dimensional regular grids such as image, video, and speech,
to high-dimensional graph-structured data, such as social networks, brain
connections, and knowledge graphs.
This evolution has led to large graph-based irregular and sparse models
that go beyond what existing deep learning frameworks are designed for.
Further, these models are not easily amenable to efficient, at scale, acceleration on parallel hardwares (e.g. GPUs).
%In this paper, we demonstrate that, by leveraging the underlying model structures, we can transform
%a graph-based neural network (GNN) into a more regular and dependent computation to take
%advantage of exsiting deep learning frameworks.
%We introduce NeuGraph, an embedded GNN processing

We introduce \pn{}, the first parallel processing framework for graph-based deep neural networks (GNNs).
\pn{} presents a new SAGA-NN model for expressing deep neural networks as vertex programs with each layer
in well-defined (\emph{Scatter, ApplyEdge, Gather, ApplyVertex}) graph operation stages.
This model not only allows GNNs to be expressed intuitively, but also facilitates the mapping to an efficient dataflow representation.
\pn{} addresses the scalability challenge transparently through automatic graph partitioning and chunk-based stream processing out of GPU core or over multiple GPUs, which carefully considers data locality, data movement, and overlapping of parallel processing and data movement.
\pn{} further achieves efficiency through highly optimized Scatter/Gather operators on GPUs despite its sparsity.
 Our evaluation shows that \pn{} scales to large real graphs that none of the existing frameworks can handle directly,
while achieving up to about 4 times speedup even at small scales over the multiple-baseline design on \tf{}.
% cannot efficiently handle
%the incurred by the nature of graph data, .
%We present GML, a GPU-based graph engine that bridges graph computation and ML.
%GML first represents the structure of ML
%model as a multiple stage of edge-centric and vertex-centric abstractions.
%Then it uses a new execution engine to accelerate the edge-centric computations, and integrates
%existing ML engines to support complex ML models in the vertex-centric computations.
%Preliminary evaluations on a single-node prototype demonstrate that the efficacy of this approach.
%GML achieves orders of magnitude speedup than PowerGraph systems, and up to 5.65x speedup than the direct implementation in TensorFlow.

%!TEX root = ../ms.tex
\section{Introduction}

Deep learning, in the form of deep neural networks
(DNNs), has been gaining popularity due to its success
in areas such as speech, vision, and natural language processing.
In these areas, the coordinates of the underlying data representation
often have a regular grid structure, which is friendly to
hardware acceleration (e.g., GPU) with
massive SIMD-style data-parallelism.
%And this achieved efficiency is one of the major factors that
%drive the success of deep learning in these areas.
There is an emerging trend in applying deep learning models on data with an irregular graph structure~\cite{defferrard2016convolutional,kipf2016semi,li2015gated,hamilton2017inductive,henaff2015deep,ggcn,googlegnn,encoding},
driven by the importance of the graph data
such as social networks, knowledge graphs, and graphs in bioinformatics and neuroscience (e.g., protein-protein interactions or neuron connections in brains),
and moving the state-of-the-art prediction results in their targeted applications (e.g., classification, embedding, and query-answering).
These graph-based neural networks (GNNs) typically apply neural network models over the features associated with vertices and edges in a graph, and propagate and aggregate the results to
produce the next-level features.

None of the existing solutions supports GNNs well.
The existing graph process engines~\cite{Pregel10,GraphLab12,gonzalez2012powergraph,powerlyra15,tux2017} often provide a Gather-Apply-Scatter (GAS)-like vertex-program model,  but incapable of expressing and supporting neural network architectures within the graph constructs. Deep learning frameworks such as TensorFlow~\cite{tensorflow16}, PyTorch~\cite{pytorch}, MxNet~\cite{mxnet15}, and CNTK~\cite{cntk14} are designed to express neural networks as dataflow graphs, but do not support a graph-propagation model directly. In addition, none of them offer the needed scalability to handle large graphs, nor do they support efficient GPU-based implementation of graph propagation operators (which translate into sparse operations).
The current lack of support has seriously limited the ability to explore the full potentials of GNNs at scale, as the combination of DNNs and large graph structures poses significant challenges at the system level.

%Although it is possible to implement GNNs using existing deep learning frameworks by treating graph structures as adjacency matrices and treating value propagations along edges in a graph as operations similar to matrix multiplications, practical issues remain when supporting such computations on large graphs. When employing GPU acceleration, the deep learning frameworks assume that the input and output tensor data of a single operator can be held in GPU device memory. However, when treating the vertex or edge data of an entire graph as a single tensor or matrix, this assumption often fails to hold. This makes these frameworks unable to support GNNs over large graphs and exploit acceleration power of GPU at the same time.
%%This limits the capability of these frameworks in supporting GNNs on large graphs and exploiting acceleration power of GPU at the same time.

In this paper, we present \pn{}, the first system to support large-scale GNNs, from an easy-to-express programming model to a scalable and efficient parallel processing engine on GPUs.
\pn{} naturally combines dataflow with a vertex-program abstraction in a new model called SAGA-NN
(Scatter-ApplyEdge-Gather-ApplyVertex with Neural Networks).
Whereas SAGA can be considered a variant of the GAS model,
the user defined functions in the SAGA-NN model allow users to express neural network computation over vertex or edge data (which are treated as tensors) by using a dataflow abstraction, rather than those designed for traditional graph processing (e.g., algorithms such as PageRank, connected component, and shortest path)

Just as with DNNs, efficient use of GPUs is critical to the performance of GNNs and is more so due to the additional challenge of handling large graph structures.
To achieve scalability beyond the physical limitation of GPUs,
\pn{} transparently partitions a graph (vertex and edge data) into chunks, and translates a GNN algorithm expressed in the SAGA-NN model into a dataflow graph with operators at the chunk granularity, through which it enables chunk-based parallel stream processing on a single or multiple GPUs.
 %the computations on entire graph data chunk-by-chunk in GPU.
 %It can also scale the computation by exploiting parallelism on multiple GPUs.
 %hides the complexity of dealing limited memory in GPUs, which is a key scalability bottleneck when mapping GNNs directly to dataflow graphs.

The efficiency of the \pn{} engine then hinges heavily on how well \pn{} manages and schedules the parallel stream processing and the implementation of the key graph propagation operators, Scatter and Gather, on GPUs.
\pn{} pays careful attention to data locality to minimize swapping data in and out of GPU memory and to maximize the reuse of data chunks while in GPU memory, while overlapping the data movement and the computation in a streaming way.
For the multi-GPU case, it uses a ring-based streaming mechanism to avoid redundant data movement from host memory by exchanging data chunks among GPUs directly.
The Scatter and Gather stages in the SAGA-NN model conduct the vertex data propagation along edges and behave as matrix multiplications on sparse structures. It is notoriously difficult to perform sparse matrix operation on data-parallel hardware like GPU. \pn{} therefore introduces special operators supported by a graph propagation engine into the dataflow graph and optimizes its execution on GPUs. Note that, unlike the traditional graph processing scenarios that other GPU-based graph engines focus on, in GNN scenarios, the mutable vertex data itself may not be accommodated in GPU device memory since the data of each vertex can be a feature vector rather than a simple scalar. Our scheme, therefore, prefers to exploit parallelism in per-vertex data access to benefit more on memory access efficiency.

We implement \pn{} through extending TensorFlow with vertex-program abstraction and custom operators for graph propagation procedure.
We demonstrate that \pn{} can scale to support a variety of GNN algorithms on large graphs containing millions of vertices with hundreds of feature dimensions and hundred millions of edges through leveraging the host memory of a single server and the computation power of GPU(s), which cannot be achieved by using existing deep learning frameworks directly. Compared with TensorFlow on small graphs that it can support with GPU, \pn{} obtains up to about 4x speedup. We also extensively evaluate the improvements caused by the multiple optimizations in \pn{} to demonstrate their effectiveness.
%In the case of a single GPU node, a direct scale-up implementation using
%the Tensorflow operators combined with data partition/swapping mechanisms could only support limited types of GNNs. 

%Compared
%against with this baseline, \pn{} can achieve up to $2.6 \times$ speedup due to overlapping the data movement
%and the computation in a streaming way and exploiting massive parallelism of
%GPU cores. 
%Our evaluation also demonstrates that \pn{} is able
%to scale nearly linearly with the number of GPU nodes due to exploiting the parallelism of multiple GPUS in a single server.

The rest of the paper is organized as follows. Section~\ref{sec:saga-nn} introduces the SAGA-NN programming abstraction. Section~\ref{sec:system} describes the components, mechanisms, and optimizations in \pn{} system. Section~\ref{sec:para} presents the ring-based streaming scheme in \pn{} to scale for multiple GPUs. Section~\ref{subsec:apps} illustrates the usage of the SAGA-NN model with applications. Section~\ref{sec:eval} discusses the implementation and evaluation of \pn{}. We discuss related work in Section~\ref{sec:related} and conclude in Section~\ref{sec:con}.

\section{\pn{} Programming Abstraction}
\label{sec:saga-nn}
\begin{figure}[t]
\centering
\includegraphics[width=150 pt]{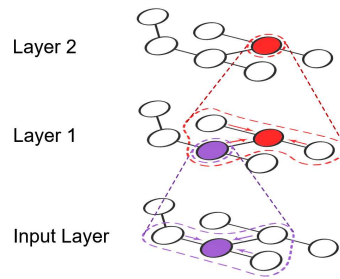}%\vspace{-2mm}
%\includegraphics[width=0.95\linewidth, trim={5cm, 7cm, 4.5cm, 0}, clip]{fig/overview_layers2_cropped}%\vspace{-2mm}
%\vspace{-2ex}
\caption{Feed-forward computation of a 2-layer GNN. }
\label{fig:overview_layers}%\vspace{-4mm}
\end{figure}

Graph-based neural network (GNN) is a general neural network architecture
defined according to a graph structure.
Each vertex or edge in the graph can be associated with a tensor data (normally a vector) as its feature or embedding.
GNN can be stacked in multiple layers, with an iterative propagation procedure conducted layer-by-layer over the same graph. In each layer, the vertex or edge features are transformed and propagated along edges and aggregated at the target vertices to produce the new features for the next layer. The transformation can be an arbitrary DNN computation. The graph may also contain a label for each vertex, each edge, or the entire graph,  for computing a loss function at the top layer. A feed-forward computation is then performed from the bottom layer to the top layer, with the back-propagation conducted reversely. Figure~\ref{fig:overview_layers} illustrates the feed-forward computation of a 2-layer GNN.

%Unlike traditional CNNs or RNNs, where the notion
%of neighborhood is given by the euclidian distance or
%the adjacent steps in the sequence,
%^For example, in regular ConvNets (or RNNs), the notion
%of neighborhood is given by the euclidian distance (or the previous step in the sequence).
%However, for graphs,
%the notion of neighborhood in GNN is given by a graph structure.
%As illustrated in Figure~\ref{fig:overview_layers}, GNN consists of multiple layers and trains a set of layer-wise aggregation functions
%where each layer aggregates the previous-layer feature information from a node's local neighborhood.
%With the recursive layer-by-layer aggregation, each
%layer-wise aggregation function actually aggregates information from a different number of hops away
%from a given node. This GNN architecture for graphs combines the
%power of neural networks and graph-based label propagation.

%Specifically,
%To ground
%the \pn{} programming abstraction in a concrete problem,
We now use the
Gated Graph ConvNet (G-GCN) algorithm~\cite{encoding,ggcn} as a concrete example.
Graph ConvNet generalizes the notion of the convolution operation,
typically applied on image datasets, to apply
to an arbitrary graph (e.g., a knowledge graph).
Gated Graph ConvNet further incorporates
the gates mechanism, so that the model can learn
which edges are more
important for the learning target.
The feed-forward computation of G-GCN at each layer is formalized recursively in Example 2.1.
%The model can be
%applied in many semi-supervised or unsupervised
%graph clustering problems such as entity
%classification in.

\begin{restatable}{example}L
Let $\mathbf{h}^{\ell}_{u}$ denote the
feature vector of a vertex $u$ at layer $\ell$, and $W ^{\ell}$, $W^{\ell}_{H}$, and $W_{C}^{\ell}$ are the weight parameters to learn.
The gated graph ConvNet algorithm recursively
defines the feature of a vertex $u$ as follows:
\begin{equation}
\begin{aligned}
\mathbf{h}^{\ell+1}_{u} = \text{ReLU} \left( W ^{\ell}\left(\sum_{v \rightarrow u} \eta_{v u}\ \odot\ \mathbf{h}^\ell_{v} \right)\right) \nonumber
\end{aligned}
\end{equation}
where $\odot$ is the element-wise multiplication and $\eta_{v u}$ (for each edge $v \rightarrow u$) acts as edge gate computed by:
\begin{equation}
\begin{aligned}
\eta_{vu} = \text{sigmoid} \left(W^{\ell}_{H}\mathbf{h}^{\ell}_{u}\ +\ W_{C}^{\ell} \mathbf{h}^\ell_{v} \right) \nonumber
\end{aligned}
\end{equation}
\end{restatable}

\begin{figure}[t]
\centering
 \begin{minipage}[t]{0.95\linewidth}
 \begin{lstlisting}[mathescape]
vertex$^{\ell + 1}$ = G-GCN(vertex$^{\ell}$)
  params p = $ [W_H^{\ell}\ W_C^{\ell}\ W^{\ell}] $
  // Passing data over edges
  edge$^{\ell}$=Scatter(vertex$^{\ell}$);
  // edge-parallel computation
  acc = ApplyEdge(edge$^{\ell}$, p)
  $\quad \eta=\text{sigmoid} (\text{p}.W_H^{\ell}\otimes \text{edge}^{\ell}.\text{src} + \text{p}.W_C^{\ell}\otimes \text{edge}^{\ell}.\text{dest})$
  $\quad$return $\eta \odot \text{edge}^{\ell}$.src
  set Gather.accumulator = sum
  accum = Gather(acc)
  // compute new vertex data
  vertex$^{\ell + 1}$ = ApplyVertex(vertex$^{\ell}$, accum, p)
  $\quad$return $\text{ReLU} \left( \text{p}.W^{\ell} \otimes \text{accum} \right)$
 return vertex$^{\ell + 1}$
\end{lstlisting}
\end{minipage}
%\vspace{-2ex}
\caption{Gated Graph ConvNet at layer $\ell$ in the SAGA-NN model, where $\otimes$ refers to matrix multiplication.}\label{fig:ggcn}%\vspace{-4mm}
\end{figure}

%, GNN a
%graph neural networks (GNNs) consist of many layers
%where each layer can be interpreted as defining a propagation function where
%the hidden state of vertices are computed recursively by the hidden states
%of their adjacent neighbors and corresponding edge features.
%of the vertices:
%\begin{equation}
%\mathbf{h}^{\ell+1}_{u}\ \leftarrow\ f^{\ell}\left(\mathbf{h}^{\ell}_{u},\ \{(\mathbf{h}^{\ell}_{v}, \mathbf{e}_{uv}): v \rightarrow u\}\right),\label{equ:gnn}
%\end{equation}
%where $\mathbf{h}^{\ell}_{u}$ is the feature vector of a vertex $u$ at layer $\ell$ and the $\{(\mathbf{h}^{\ell}_{v}, \mathbf{e}_{uv}): v \rightarrow u\}$ denotes the set of the feature vectors of the vertices neighboring $u$ and the corresponding edge feature.
%Thus, GNNs actually parameterize an
%unfolding of the iterative message propagation procedure.

\vspace{-2ex}

\subsection{SAGA-NN Model}
\label{sec:saga}

To express the recursive computation at each layer of a GNN,
which contains both graph propagation and DNN computation,
we propose a SAGA-NN
(Scatter-ApplyEdge-Gather-ApplyVertex with Neural Networks)
vertex-program abstraction.
SAGA-NN defines four stages of a feed-forward computation in each layer of a GNN: \emph{Scatter}, \emph{ApplyEdge}, \emph{Gather}, and \emph{ApplyVertex}.

%\label{subsec:func_ops}
%We observe that a GNN algorithm essentially parameterizes an
%unfolding of an iterative message propagation procedure, which consists of an edge parallel
%operation followed by a commutative associative
%aggregation for each vertex, and finally a vertex parallel
%operation on the corresponding aggregates.
%For example, in G-GCN, the edge parallel
%operation computes $\eta_{v u}\ \odot\ \mathbf{h}^\ell_{v}$, followed by
%the sum aggregation $\left(\sum_{v \rightarrow u} \eta_{v u}\ \odot\ \mathbf{h}^\ell_{v} \right)$, and the vertex parallel
%operation computes the representation of each vertex.
%Both the edge and vertex parallel
%operations can contain learnable parameters.

\begin{figure}
	\centering
	\includegraphics[width=0.95\linewidth]{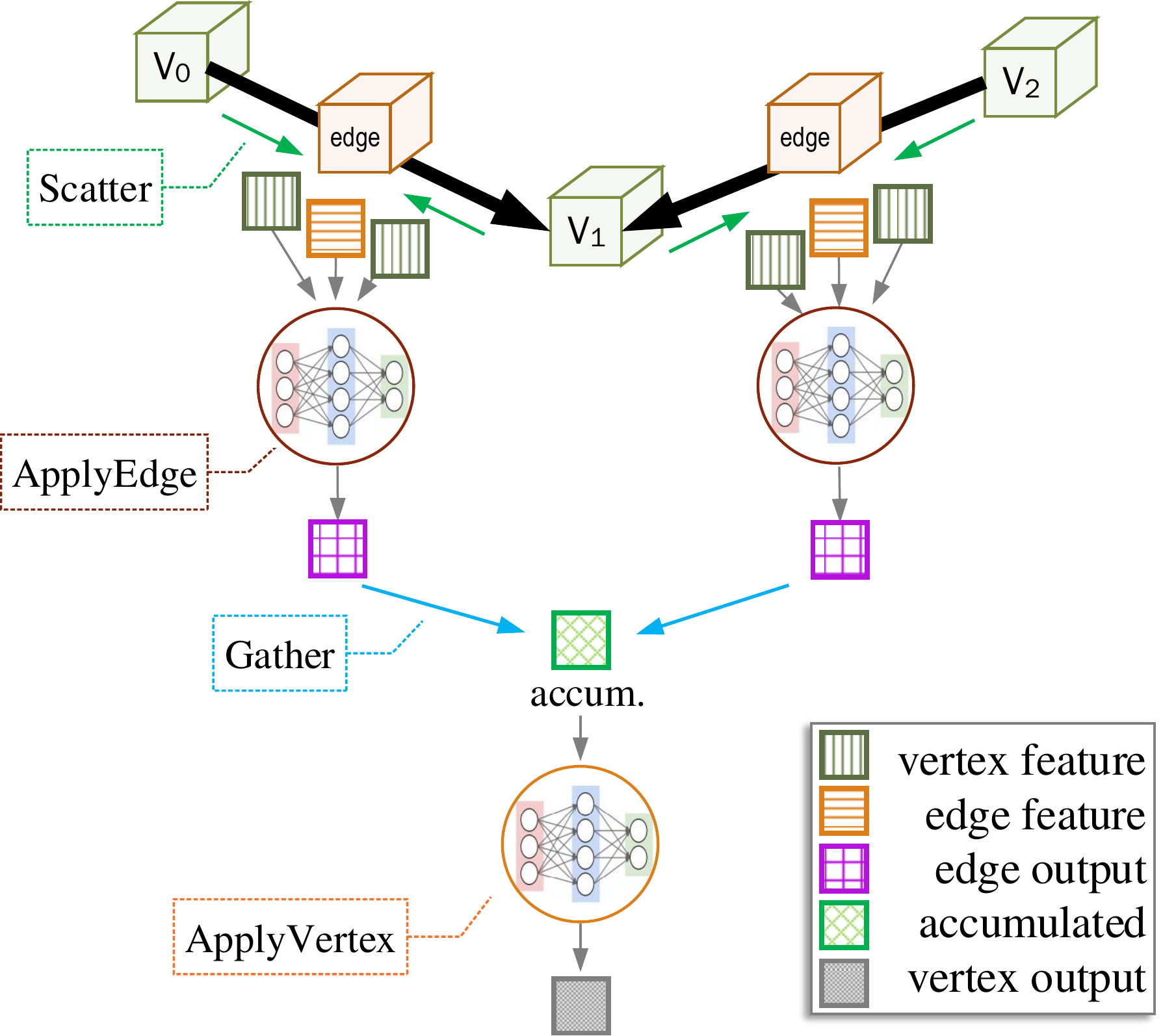}
%	\vspace{-1ex}
	\caption{SAGA-NN Stages for each layer of GNN.}
	\label{fig:sagacropped}
\end{figure}

%Based on this common pattern, we propose a SAGA-NN
%(Scatter-ApplyEdge-Gather-ApplyVertex with Neural Networks) abstraction to
%declare the layer-wise propagation rule. The SAGA-NN model consists of three main parts: the data
%abstractions, the user defined functions (UDFs) with neural networks, and the message propagation operations.
%To separate message passing and ease the programming of GNN, \pn{} abstracts the data of a node into
%\texttt{vertex} (e.g., node representation $h_u$ in G-GCN), and the data of source node and destination node of an edge into \texttt{edge}=[\texttt{src} \texttt{dest}] (e.g., a pair of node representations [$h_u,\ h_v]$ in G-GCN).

%SAGA-NN abstracts the data associated with vertex into the class \texttt{vertex} which is a tensor type as the representation of the vertex (e.g., $h_u$ in G-GCN).

SAGA-NN provides two user-defined functions (UDFs) for \emph{ApplyEdge} and \emph{ApplyVertex}, respectively, to declare the neural network computations on edges and vertices.
The \textbf{ApplyEdge} function defines
the computation  on each edge, which takes \texttt{edge} and \texttt{p} as inputs, where \texttt{edge} is the abstraction of edge data and \texttt{p} contains the learnable parameters of the GNN model.
Each \texttt{edge} is a tuple of tensors [\texttt{src}, \texttt{dest}, \texttt{data}] representing the data of the source and destination vertices connected by the edge, as well as the edge associated data; e.g., edge weight.
This function can be used to apply a neural network model on \texttt{edge} and \texttt{p}, and output an intermediate tensor value associated with the edge.
%edge data abstraction \texttt{edge}, edge static feature $\mathit{ef}$ (e.g., edge weight) and learnable parameters $\mathit{params}$
%and outputs a partial aggregated message $\mathit{acc}$.
The \textbf{ApplyVertex} function defines
the computation on a vertex, which takes as input a vertex tensor data \texttt{vertex}, the
vertex aggregation \texttt{accum} and learnable parameters \texttt{p}, and returns the new vertex data through applying a neural network model.
The SAGA-NN abstraction builds on a dataflow framework, so users can symbolically define the dataflow graphs in UDFs by connecting
mathematical operations (e.g., add, tanh, sigmoid, matmul) provided by the underlying framework.

The other two stages, \emph{Scatter} and \emph{Gather}, perform data propagation and prepare data collections for inputs of \emph{ApplyEdge} and \emph{ApplyVertex}. They are triggered and conducted by the system implicitly, and do not require users to provide explicit UDFs.
%Besides UDFs encoding computation, SAGA-NN provides two key internal operations
%that manages data dissemination and collection before ApplyEdge and ApplyVertex functions.
%The \textbf{Scatter} operation passes the
%vertex data onto the adjacent edges, whereas the \textbf{Gather} operation
%propagates the intermediate representations of edges produced by \textbf{ApplyEdge} and aggregates them at the destination vertices.
%passes $acc$ summed to the accumulative value for the destination node.
%The general aggregation method \textbf{sum} is commutative and associative. \pn{} provides a series of default methods, e.g., max, sum, and concatenation, which can be chosen through setting \textbf{Gather}.\texttt{accumulator}.
Figure~\ref{fig:ggcn} illustrates the description of G-GCN (at layer $l$) in the SAGA-NN model.

\subsection{Execution Semantics}

For each GNN layer, the four-stage execution flow of a feed-forward computation is illustrated in Figure~\ref{fig:sagacropped}. It starts from the \emph{Scatter} stage, where the tensor data \texttt{vertex} of each vertex is passed onto the adjacent edges to construct edge data \texttt{edge}, containing both the source and destination vertex data. The subsequent \emph{ApplyEdge} stage then invokes a parallel computation defined by the \textbf{ApplyEdge} function on the edge data to produce an intermediate tensor value for each edge as its outputs. The \emph{Gather} stage then propagates those outputs along the edges and aggregates them at the destination vertices through commutative and associative accumulate operations. Finally, the \emph{ApplyVertex} stage executes the computation defined in \textbf{ApplyVertex} function on all vertices to produce updated vertex data for the next layer.

A GNN training process also involves a backward computation phase in each stage of a layer $\ell$ for back-propagation of gradients. The backward computations are invoked in the reverse order of the feed-forward computation: it starts from the \emph{backward-VertexApply} stage,
which takes the gradient $\nabla \text{\texttt{vertex}}^{\ell+1}$ passed from layer $\ell + 1$ as input to update the corresponding learnable parameters on vertices, and outputs the gradient ${\nabla}$\texttt{accum} and the partial vertex gradient $\nabla\text{\texttt{vertex}}^{\ell}_p$. Then the \emph{backward-Gather} stage takes ${\nabla}$\texttt{accum} as input and computes its output gradient $\nabla \texttt{acc}$ for each edge based on the accumulate function used in \emph{Gather}. The subsequent \emph{backward-ApplyEdge} stage then takes $\nabla \texttt{acc}$ as input to update the learnable parameters relevant to edge, and outputs the partial vertex gradients $\nabla\text{\texttt{vertex}}^{\ell}_p$ for both the source and destination vertices. Finally, the \emph{backward-Scatter} stage accumulates all the partial gradients for a vertex to generate the final $\nabla \text{\texttt{vertex}}^{\ell}$ and pass to layer $\ell-1$.

\pn{} can automatically generate the corresponding back-propagation execution for each layer of GNN defined in the SAGA-NN model, because the UDFs for \textbf{ApplyEdge} and \textbf{ApplyVertex} are expressed as dataflow computations over regular tensors and can therefore leverage auto-differentiation provided by the deep learning frameworks.
We choose not to expose UDFs for \textbf{Scatter} and \textbf{Gather},
because these UDFs, if provided, are highly coupled with the propagation procedure, whose computations flow through the irregular graph structure and are hard to be expressed as dataflow that \pn{} optimizes---users would have to implement the corresponding derivative functions of the UDFs, a serious burden.
%over tensor objects at an appropriate granularity, as required by efficient enough execution.
%Consequently, users cannot leverage high-level dataflow abstraction to express these UDFs and have to implement the corresponding derivative functions of the UDFs by themselves, which is a significant burden. 
Following the same principle, \pn{} also avoids to expose user-defined aggregation methods. 
It provides a set of default ones instead, including max, sum, and concatenation, which can be chosen by setting
\textbf{Gather}.\texttt{accumulator} (as shown in Figure~\ref{fig:ggcn}).
%We instead provide a set of default ones including max, sum, and concatenation for defining
% \textbf{Gather}.\texttt{accumulator} (as shown in Figure~\ref{fig:ggcn}).

By carefully combining  dataflow with the vertex-program abstraction, SAGA-NN inherits benefits from both. The dataflow abstraction makes it easy to express neural network architectures in GNNs and leverage auto-differentiation.
The vertex-program in SAGA-NN allows users to express computations naturally by thinking locally as a vertex
%, which is very natural for graph-structured data,
and captures common patterns in GNNs as well-defined stages, thereby enabling graph-related optimization (e.g., efficient graph propagation procedures) and helping produce streaming-based dataflow graph with an optimized scheduling strategy.

\section{\pn{} System}
\label{sec:system}
%\section{NeuGraph System}

\pn{} provides a combination of the dataflow and vertex-program abstractions as the user interface. Underneaththis abstraction, \pn{} mainly consists of 1) a front-end  that translates the algorithm implemented in SAGA-NN model into a chunk-granularity dataflow graph to enable GNN computation on large graph in GPU; 2) an optimization layer that produces a scheduling strategy for minimizing data movement between host and GPU device memory, and recognizes opportunities to fuse operations and remove redundant computations; 3) a set of efficient propagation operation kernels that support streaming-based processing to overlap data movement and computation in GPU; 4) a dataflow execution runtime.
Because \pn{} largely leverages existing dataflow-based deep learning frameworks for the dataflow execution runtime,
we focus on the design of the first three in this section as they are the main contributions of \pn{}.

\begin{figure*}[t]
    \centering
    \includegraphics[width=0.99\linewidth]{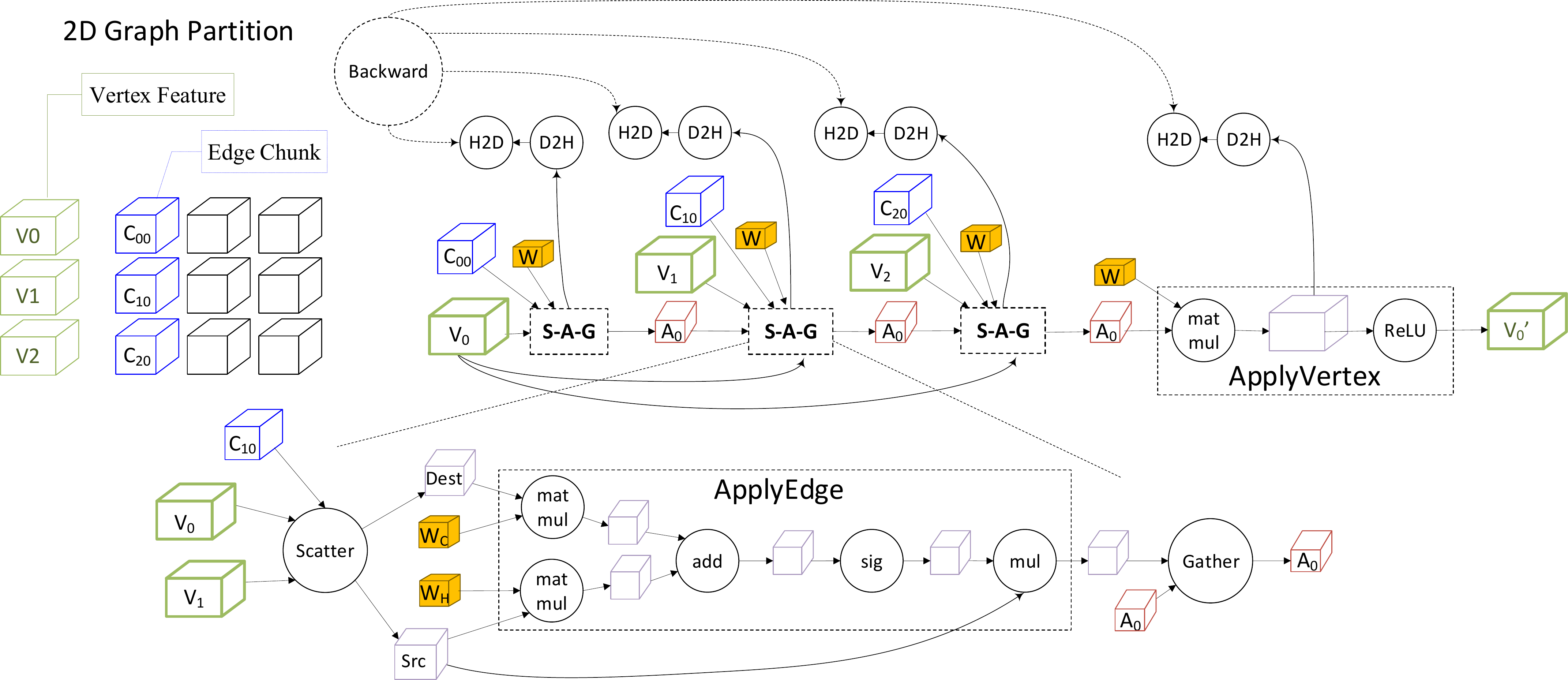}
    \vspace{-2ex}
    \caption{Chunk-based dataflow graph for a destination interval $V_0$ at a G-GCN layer.
    The swap-out of output tensors of operations
    from the SAG phase when connecting to D2H is hidden in the SAG sub-graph for a clear visualization.}\label{fig:partition_dataflow}%\vspace{-4mm}
\end{figure*}

\subsection{Chunk-Based Streaming Dataflow}
\label{sec:system:dataflow}

When exploiting computation power of GPU, existing deep learning frameworks assume that the input and output tensor data of a single operator in the dataflow graph can be fully held in GPU device memory. However, when treating the vertex or edge data of an entire graph as a single tensor or matrix, this assumption often fails to hold for large graphs, which limits the scale of the graph that the system can support efficiently. To address this problem, \pn{} splits the vertex and edge data of the graph into small chunks that can fit into GPU device memory and constructs a dataflow graph with operators processing computations at the chunk granularity.

\pn{} splits vertex and edge data into chunks through a 2D partitioning to tile the adjacency matrix representing the edges in the graph, with vertex data split into the corresponding equally-sized disjoint vertex id intervals. This way, edges in each edge chunk connect the vertices in two vertex chunks as sources and destinations, respectively. As illustrated in Figure~\ref{fig:partition_dataflow}, edge chunk $C_{ij}$ contains edges connecting source vertices in vertex chunk $V_i$ and destination vertices in vertex chunk $V_j$. In each edge chunk, for feed-forward computation, edges are laid out in a compressed sparse column (CSC) format with edges clustered by destination vertex id, whereas for back-propagation computation, edges are arranged in compressed sparse row (CSR) format. \pn{} also makes a best effort to re-encode vertex ids to equalize the numbers of edges in edge chunks for balanced chunk-granularity computation.

By splitting data into chunks, \pn{} is able to generate a dataflow graph with operators operating on data chunks that fit in GPU memory.
%, so that every operator can run in GPU efficiently.
In the dataflow graph, the outputs of the operators in the \emph{Scatter} stage are a grid of edge data chunks with edge data the tuple [\texttt{src},\texttt{dest},\texttt{data}] (in Section~\ref{sec:saga}). Each edge data chunk is then processed by the operators in the \emph{ApplyEdge} stage to produce another grid of edge data chunks with edge data \texttt{acc} (as in Figure~\ref{fig:ggcn}). The operators in the \emph{Gather} stage then accumulate these edge data chunks
%outputed by \emph{ApplyEdge} stage
based on the destination vertices to generate the corresponding vertex data chunks as the inputs for the \emph{ApplyVertex} stage that follows.

A naive scheduling strategy for this dataflow graph is to execute the operators stage-by-stage. However, since the size of the output data chunk grids of a stage can be large and may not be accommodated in the GPU device memory, they will be swapped out to host memory after the completion of the current stage and before the beginning of the next, thereby losing the opportunity for the operators in the next stage to reuse the output data chunks that already reside in GPU device memory in the current stage.
%by the current stage operators.
For example, when the \emph{Gather} stage outputs a vertex chunk with the accumulated vertex data, the scheduler may choose to schedule the operators in the same \emph{Gather} stage to produce a next vertex chunk, or, it may immediately schedule the operators in the following \emph{ApplyVertex} stage to consume the current vertex chunk since it is already in the GPU device memory.

\pn{} therefore adopts a scheduling strategy for every GNN layer as illustrated in Figure~\ref{fig:partition_dataflow}. For each output vertex chunk of the \emph{Gather} stage (e.g., $A_0$ in the figure) that maintains the accumulated values for the destination vertices in the chunk, it may be used as input and output by multiple operators in the same \emph{Gather} stage for the different source vertex chunks (e.g., $V_0$, $V_1$, and $V_2$). The scheduler tries to hold it in GPU device memory to keep reusing it for those \emph{Gather} operators until it is consumed by the operators in the following \emph{ApplyVertex} stage. As in Figure~\ref{fig:partition_dataflow}, \pn{} continuously executes operators in the \emph{Scatter}, \emph{ApplyEdge}, and \emph{Gather} (S-A-G) stages and repeats this pattern for $V_0$, $V_1$, and $V_2$ to produce the final $A_0$, followed by the corresponding operators in the \emph{ApplyVertex} stage that consumes $A_0$. It then schedules the operators to produce the next output vertex chunk of the \emph{Gather} stage.

\pn{} employs explicit device-to-host (D2H) and host-to-device (H2D) operators to conduct the data swap between host and GPU device memory. During a training process, the intermediate feature data (e.g., the result of matrix multiplication in the \emph{ApplyEdge} stage as in Figure~\ref{fig:partition_dataflow}) relevant to vertex or edge chunks may be used in back-propagation. They may be swapped out to host memory during feed-forward computation and swapped back in during back-propagation.

Employing explicit data movement operators enables streaming-based scheduling, to overlap data swap with other computations in operators that are independent of the data transferred. For example, at the beginning of the execution of each GNN layer, the operators in the \emph{Scatter} stage can overlap with their corresponding H2D operators; i.e., the scatter operation on the current vertex chunk can overlap with the operator that loads the next vertex chunk from host memory to GPU device memory.

\begin{figure*}[t]
    \centering
    \includegraphics[width=0.8\linewidth]{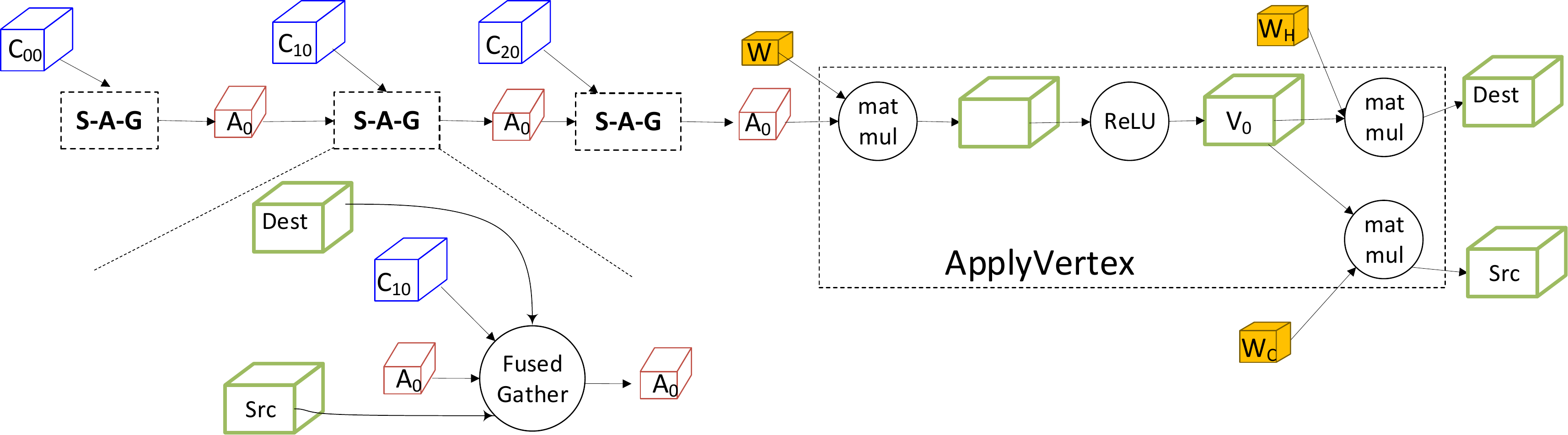}
    \vspace{-2ex}
    \caption{Optimized layer-wise dataflow graph in G-GCN.}
    \label{fig:optimized_dataflow}
    \vspace{-2ex}
\end{figure*}

%At each layer, the produced dataflow subgraphs for different vertex intervals can be executed iteratively on one GPU or to be executed in parallel on multiple GPUs.

%\para{Data Swapping.}During the training process, the intermediate feature data generated for each chunk are not de-allocated as they will be reused in backward pass.  As a result, these data gradually accumulate until the no further chunks can be loaded.
%To address this problem, \pn{} also explicitly inserts device-to-host (D2H) and host-to-device (H2D) copy operations to specify the data swapping between the GPU and CPU devices.
%For layer-wise dataflow subgraph, \pn{} attempts to swaps all the intermediate feature data generated in each phase to make the chunk size as large as possible, so that the data movement can be minimized (see Figure~\ref{fig:partition_dataflow}).
%To do so, \pn{}
%rewrites the dataflow graph by inserting D2H and H2D operations between the intermediate feature data and the backward operations reusing these data.
%So the intermediate feature data can be released immediately after it is transferred to CPU memory and not used by other operations in the froward pass.
%Notice the elementwise operation can update data locally, so \pn{} only swaps out the output tensors of non-elementwise operations.
%Notice the parameters are shared by edges and vertices at each layer, so \pn{} swaps out layer-wise parameter data onto CPU device at the end of layer-wise subgraph.

\begin{figure}[t]
    \centering
    \includegraphics[width=0.98\linewidth]{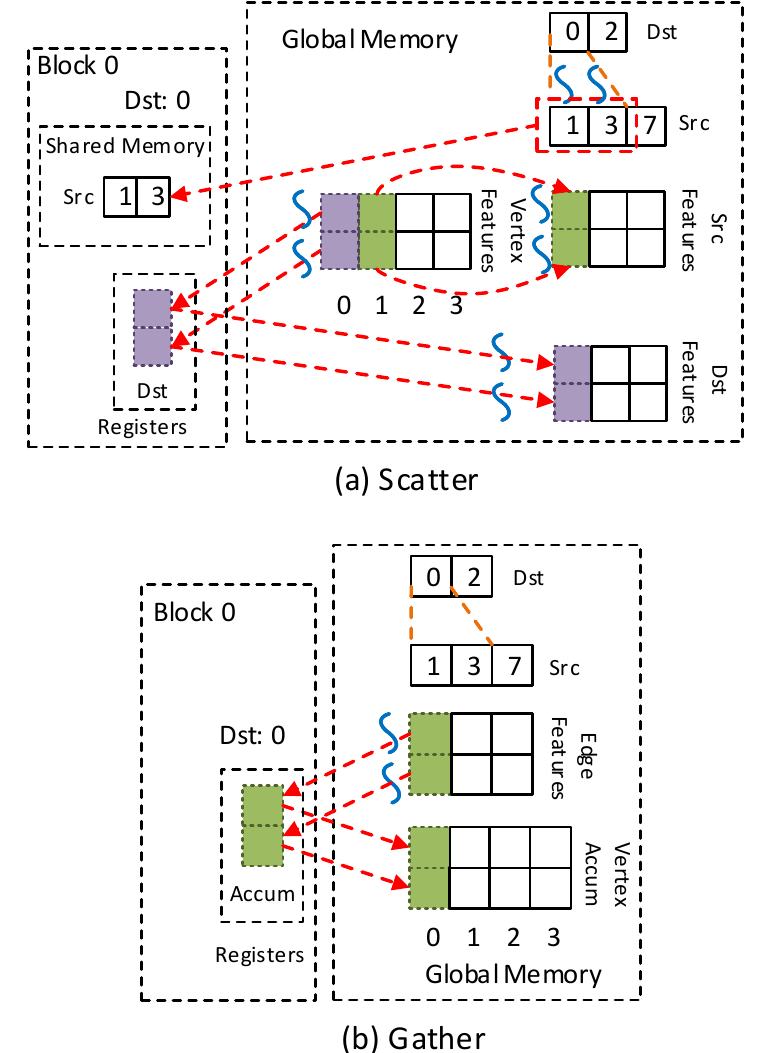}
    \vspace{-1ex}
    \caption{Propagation kernels.}
    \label{fig:optimized_kernel}
    \vspace{-2ex}
\end{figure}

\subsection{Dataflow Graph Optimization}\label{subsec:opt}
\pn{} further optimizes a generated dataflow graph to remove redundant computations or fuse operations by considering the semantics of the SAGA-NN model. Considering the matrix multiplication operations in the \emph{ApplyEdge} stage as in Figure~\ref{fig:partition_dataflow}, they perform computations between the vertex data that are scattered to the edges and the learnable parameter $W_C$ or $W_H$ that is shared by all the edges. Because a vertex may have multiple adjacent edges that the vertex data can be scattered to, such same multiplication for a vertex can be conducted multiple times and redundant. \pn{} therefore moves the computations that are only related to source or destination vertices out of the \emph{ApplyEdge} stage of the current layer to the \emph{ApplyVertex} stage of the previous layer. Figure~\ref{fig:optimized_dataflow} shows the optimized dataflow graph with the matrix multiplications moved into the \emph{ApplyVertex} stage.

\pn{} supports operator fusion as another optimization. In some GNN algorithms, the \textbf{ApplyEdge} function does not perform complex neural network computation, but only element-wise operations, such as +, -, $\times$, $\div$, tanh, sigmoid, ReLU, etc. In this case, the \emph{Scatter}, \emph{ApplyEdge}, \emph{Gather} stages can be reduced to a single propagation procedure using a special customized operator. \pn{} automatically detects this case and replaces the subgraph of these three stages with the special operator. As shown in Figure~\ref{fig:optimized_dataflow}, for G-GCN, after the matrix multiplications are moved to \emph{ApplyVertex} stage, the SAG phase for each chunk contains only element-wise operations (e.g., + and sigmoid), so the whole SAG phase is replaced with a \emph{fused-gather} operation.

\subsection{Propagation Kernels on GPUs}\label{subsec:opkernel}
\emph{Scatter} and \emph{Gather} are the two key stages to handle the propagation procedures over the often sparse edge structure of a graph. 
To support both stages efficiently on GPUs, 
\pn{} provides customized scatter/gather operation kernels optimized for GPU execution. The design carefully considers the data structure layout to allow the kernels to better leverage GPU memory hierarchy (i.e., shared memory and register) and massive parallelism. Even with a sparse graph structure, which often leads to random access on vertex data, unlike traditional graph processing scenarios, in most GNN algorithms, the data of each vertex is a dense vector rather than a scalar, we therefore prefer to exploit GPU parallelism in per-vertex data access with a best effort.

\para{Scatter Kernel.}
The scatter operator passes vertex feature data, from both source and destination, to edges. In feed-forward computation, edges are arranged in a CSC format. We therefore organize the incoming edges of a single destination vertex as in a group and assign a thread block to process them. For a vertex with a large in-degree, we split the edge set into consecutive subgroups for multiple thread blocks. In each thread block, the process of scattering source vertex data consists of two phases, as illustrated in Figure~\ref{fig:optimized_kernel}(a). First, the threads fetch the source vertex ids from the edges in the edge group stored in contiguous address space, and cache them in the shared memory. Second, the threads copy the vertex data to the edge data storage in parallel along the dimensions of vertex feature vector. This way, in most cases, an instruction in a thread-warp may have highly efficient coalesced memory access. The destination vertex data can be accessed in a similar way. Because destination data is shared by multiple edges, it can be cached in register after being accessed once. In back-propagation, a similar process is conducted on the CSR edge format.

%In GNN scenarios, the vertex data is often a feature vector rather than a simple scalar, thus we choose to exploit GPU parallelism in per-vertex data access, i.e., each thread processes one dimension data of the feature vector, to benefit memory access efficiency. In \pn{}, we organize all the incoming edges of a single destination vertex as a edge group, which is processed by a single thread block. Multiple edge groups from different destination vertices are concurrently processed on many streaming-multiprocessors (SM).

%To reduce the non-coalesced memory accesses of the CSC graph representation,
%for each edge group, we first cache its edge index data into shared memory, where each thread loads one edge's index.
%To scatter the destination vertex's feature to edges, consecutive threads of a thread block first loads consecutive elements of the destination feature vector from global memory to register, and then write it to each edge with vectorized memory access.
%For source vertices, we simply use the same threading model as above to copy each vertex feature to the corresponding edges, as shown in Figure~\ref{fig:optimized_kernel}(a).

%Note that, if the edge size of a group from single destination vertex is larger than the size of thread block, we will split the edge set into consecutive subgroups, and process them one by one, until all the edges are processed.

\para{Gather Kernel.}
The gather operator collects the feature vectors of edges to the destination vertices and reduces them into a single vector for each destination vertex through a user-provided accumulate function. We employ a similar principle of exploiting parallelism as for the scatter operator. The process is illustrated in Figure~\ref{fig:optimized_kernel}(b). A block of threads first cooperatively enumerate the edge group, and accumulate the features of each edge into a temporary vector in register, and then write the result back to the corresponding destination vertex.

\section{Parallel Processing with Multiple GPUs}
\label{sec:para}
%To scale out the computation in multiple GPUs settings, a vanilla solution would be running streaming-based propagation from Section~\ref{sec:system} on each GPU concurrently. Then one GPU will take care of a destination vertex chunk each time, independently.
\pn{} further exploits the parallelism of multiple GPUs in a single server.
%, by leveraging the characteristic of hardware architecture.
 Figure~\ref{fig:mgpu-arch} shows the interconnection architecture of a typical 8-GPU server, where GPUs are connected to CPU/DRAM (host memory) via a multi-level PCI-Express (PCIe) interface hierarchy.

%\para{Multi-GPU Architecture}
\subsection{Multi-GPU Architecture}
\label{subsec:mgpu-arch}

\begin{figure}[t]
	\centering
	\includegraphics[width=0.95\linewidth]{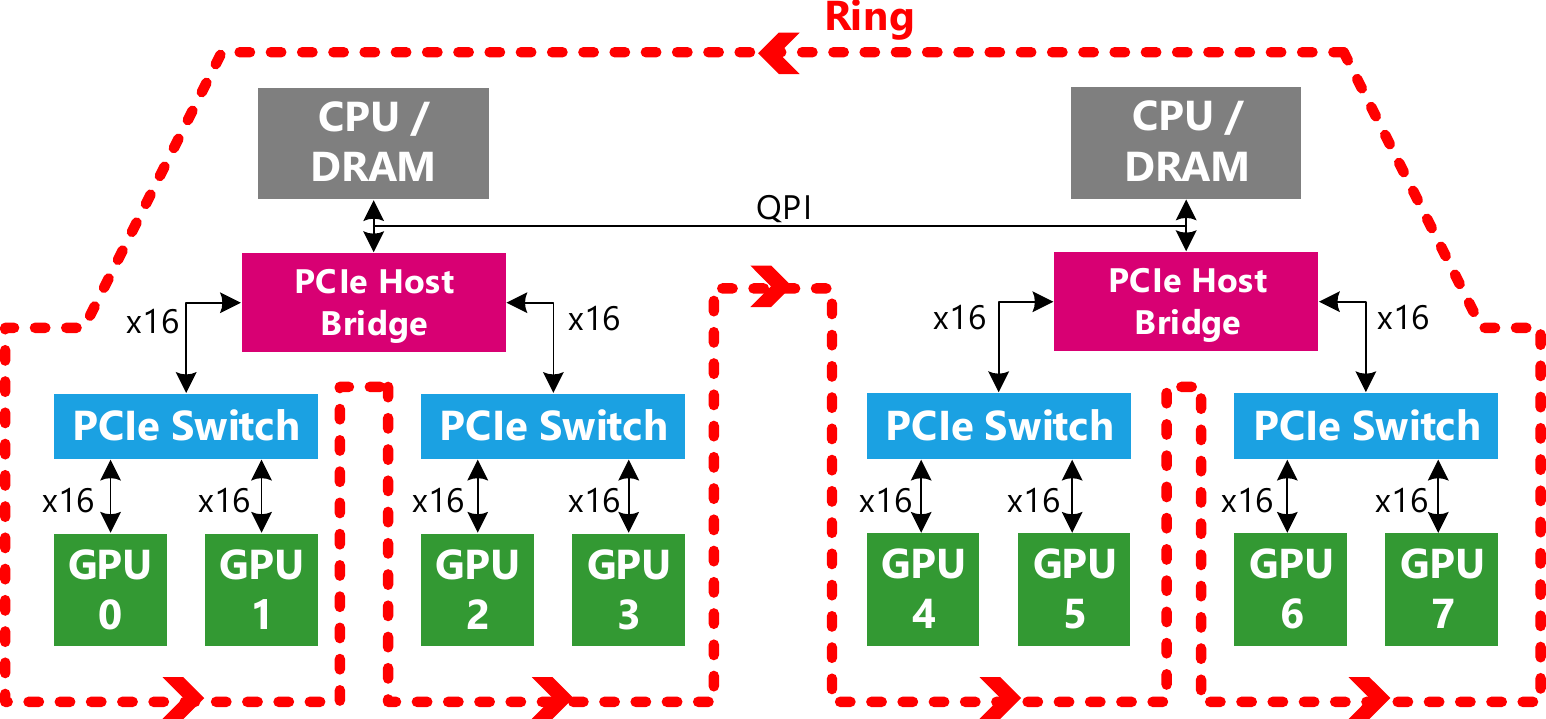}
	\vspace{-2ex}
	\caption{Multi-GPU architecture}
	\label{fig:mgpu-arch}
	\vspace{-1ex}
\end{figure}

In a multi-GPU system, the GPU interconnection has a hierarchical structure, leading to multiple levels of locality when transferring data across different GPUs or between host and GPU device memory. For example, as shown in Figure~\ref{fig:mgpu-arch}, the communication between GPU~0 and GPU~1 achieves the highest performance as they are attached to the same PCIe switch. GPU~0 needs to go through two PCIe switches and one PCIe host bridge when communicating with device like GPU~2 or 3, therefore introducing longer latency and lower throughput. GPU~0 cannot conduct P2P access when communicating with GPUs located in other NUMA nodes (e.g., GPU~4 or 7), and hence performs even worse. 

In addition, the upper level link bandwidth is shared by the GPU devices under the same root. For example, GPU~0 and GPU~1 share the same communication bandwidth between the PCIe host bridge and the PCIe switch that they are rooted at.
As a result, their communications with GPU devices under other PCIe switches may interfere with each other when conducted concurrently, leaving their local PCIe bandwidth under-utilized. This makes the root level links more likely become the bottleneck in parallel computation.
We therefore carefully design the streaming scheme by considering the locality characteristics of the multi-GPU system to achieve higher parallelism and better performance.

%\para{Locality in multi-GPU architecture}
%In a multi-GPU system,
%the device--device
%communication performance
% between two GPUs
%vary through different link paths.
%For example, in Figure~\ref{fig:mgpu-arch}, GPU~0 gets the highest device--device speed when communicating with a device that attached to the same PCIe switch, i.e., GPU~1;
%, with link path traversing only a single PCIe switch;
%GPU~0 needs to go through 2 PCIe switches and a PCIe host bridge when communicating with device like GPU~2--3, therefore in slower speed;
%GPU~0 cannot perform P2P access when communicating with GPUs locate on other NUMA nodes, e.g., GPU~4--7, therefore introducing about twice the latency.
%All the links in Figure~\ref{fig:mgpu-arch} build a tree network.
%However we should notice that the links in upper level cannot match the bandwidth of its children links. For example, a PCIe switch has a x16 upper link, with children links doubled total bandwidth.
%single direction bandwidth of these links are similar, 15.8 GB/s for PCIe 3.0 x16~\cite{pcie3} and 19.2--25.6 GB/s
%per link pair
%for QPI~\cite{intelQPI}.
%So the root level links are more likely to become a bottleneck, especially when there are multiple devices communicating sharing this link.

%local bandwidth is high
%\para{Multi-level Locality}
%We already know that,
%Besides the conclusion that ``the closer 2 devices are, the better communication performance we can get'',
%a similar rule also applies to higher level of the tree network, for example in PCIe switch level communication --
%A PCIe switch can have better communication performance with a switch in the same NUMA node.
%the closer 2 PCIe switches are, the better communication performance we can get.
%we should better have the communication taken locally as much as possible.
%avoid long-step path, avoid root path

%Besides, both PCIe and QPI links support duplex~\cite{pcie3,intelQPI}, which means we can only fully utilize their bandwidth when communicating on both directions at the same time.
%These characteristics suggest that in the design of communication, we should consider locally.
%To better utilize the bandwidth of all the links in such multi-GPU architecture for better performance, we propose a ring-based multi-stream parallel processing model.

\subsection{Ring-Based Parallel Streaming}

%\para{Locality-aware Access}
Consider a layer of GNN computation, a vanilla solution to exploit multi-GPU parallelism is to run a dataflow subgraph (as shown in Figure~\ref{fig:partition_dataflow}), which produces an output vertex chunk of the \emph{ApplyVertex} stage, in a single GPU, and let multiple GPUs execute different sets of such subgraphs. The process of this subgraph essentially outputs one vertex chunk (e.g., $V'_{0}$) and takes a set of vertex chunks containing all vertices (e.g., $V_0$, $V_1$, and $V_2$) as input. When running these subgraphs in multiple GPUs at the same time, the same set of vertex chunks are loaded from host memory into the device memory of these GPUs. This makes data transfer bottlenecked at the root level PCIe links that are shared by all the GPUs. Therefore, we design a ring-based streaming scheme to allow GPUs to reuse the data chunks that are already loaded from host memory by exchanging them directly. We organize the data-sharing path among the multiple GPUs as a ring, which is illustrated as the circle with red dot line in Figure~\ref{fig:mgpu-arch}. Because both PCIe and QPI links support duplex~\cite{pcie3,intelQPI}, the simultaneous data transfers on the ring do not interfer with each other on bandwidth. With this scheme, each vertex chunk is loaded from host memory to enter the ring and passed to all the GPUs on the ring in order.

\begin{figure}[t]
	\centering
	\includegraphics[width=0.95\linewidth]{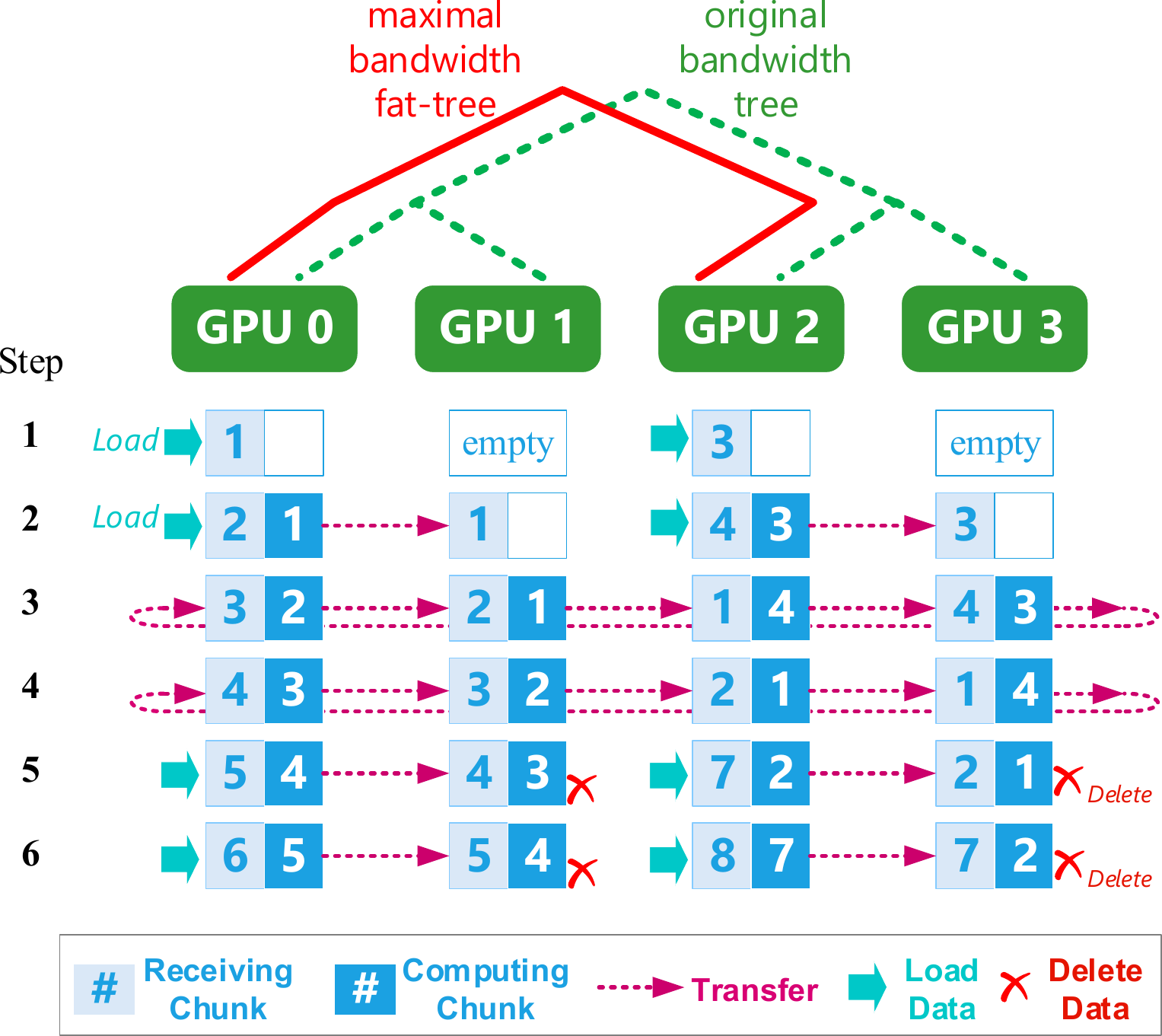}
	\caption{Ring-based streaming in a 4-GPU setting.}
%	\vspace{-2ex}
	\label{fig:load-ring}
%	\vspace{-1ex}
\end{figure}

\para{Scheduling in Ring-Based Streaming.}
%All Forward
In ring-based streaming, a GPU needs to take the following two actions:
1) loading a data chunk from the host memory or from the device memory of the previous GPU in the ring; 2) performing local computations. In order to enable the overlap between the two actions, \pn{} introduces an explicit device-to-device (D2D) operator and employs a coordinated scheduling as illustrated in Figure~\ref{fig:load-ring}.
%; and (3) transferring input data to the next GPU if still useful to next, or delete if used by entire ring.
%To make sure the ring going forward, GPUs also need to coordinate with 
% one another 
%its both sides neighbor in ring.
%As shown in Figure~\ref{fig:load-ring}, we use an 4-GPU example to explain how ring-based stream works in \pn{}.
%Each GPU has buffers for receiving data and for computation, holding data called \emph{Receiving chunk} and \emph{Computing chunk} respectively.
%Our system supports multiple buffer slices for multiple data chunks, but here we use single buffer chunk for simplicity.
%The buffer for computation can also be read by next neighbor GPU in ring.
%Before starting a \emph{ring-forward operation}, we first need to load data into device memory, as initially the data locating in host memory.

In step~1, only GPU~0 and GPU~2 load vertex data chunks 1 and 3 from the host memory, respectively.
After loading, in step~2, GPU~0 and GPU~2 start computations based on chunks 1 and 3, while loading chunks 2 and 4, respectively. 
At the same time, GPU~1 and GPU~3 start fetching chunks 1 and 3 from GPU~0 and GPU~2.
In step~3 and step~4, they enter the whole-ring forward mode where there is no data loaded from the host memory. 
%With buffers fulfilled, each GPU receives a new data chunk from its previous GPU while performing computation on local computation chunk(s).
In step~5, unable to get input vertex data from the ring any more, GPU~0 and GPU~2 read data chunk 5 and 7, respectively, from the host memory.
Additionally, GPU~1 and GPU~3 drop data chunks 3 and 1 after processing them locally because the chunks have already been consumed by all the GPUs in the ring.
All the GPUs in step~6 take the same actions as step~5, except that the data chunk numbers are shifted. And the process enters the whole-ring forward mode again after that step. The whole process continus in such a pipelining fashion until all vertex data chunks have been loaded and processed.
%If compare step 5,6 with 1,2, we can notice that they are quite similar. This ``bumpy'' pattern makes the data-loading and data-computing in a ring not only overlapped, but also pipelined. 

This scheme leaves GPU~1 and GPU~3 idle in step~1 because it already makes full use of the system bandwidth. More concurrent loadings can make things worse.
%load through a ``maximal fat tree" to devices
For example, if all the GPUs load different vertex data chunks from the host memory simultaneously, throughput will be limited by root links such as the upper link of PCIe switch in Figure~\ref{fig:mgpu-arch}.
In this case, GPU~0 and GPU~1 will get only half the bandwidth due to the shared upper link. Suppose a single GPU needs time $t_{0}$ to transfer a data chunk into GPU device memory,
% and $t_{1}$ for processing.
two concurrent transfers will take $2 \cdot t_{0}$
%$2 \cdot t_{0} + t_{1} + 2 \cdot t_{0}$
before starting computation.
%, as each device can only use half bandwidth of the shared upper link.
%On the contrary, GPU~0 transferring then GPU~1 only takes $t_{0}$ (as exclusive upper link) for loading data chunk, and can then begin processing.
% while having GPU~1 transferring, with total time $t_{0} + max((t_{0} + t_{1}), t_{1})$.
%The latter solution benefits in ring processing.
Actually, such longer transfer latency will happen each time the ring loads new data chunks from the host memory, which leads to extra stalls in the entire processing pipeline.

%In Figure~\ref{fig:load-ring}, we give an example of how a 4-GPU ring starts loading input data and processing.
%We can see that only GPU 0 and GPU 2 loading data from host memory in first 2 steps will be enough to boot up the ring processing.
%Besides, loading data for next round of ring can be pipelined with previous round, as show in step 5 and 6.

%At the end the ring-based streaming process, GPUs finish their work at different steps, just like how they start. Similar to data loading, the device-to-host bandwidth does not allow all the GPUs transfer in full-speed simultaneously.
%And ``first finished first output to host memory'' will naturally pipeline the transferring.

In addition, analyzing the multi-GPU architecture can help decide which GPU(s) should perform data loading in ring-based streaming.
With host memory as root, GPUs as leaves, and data links (with bandwidth) as edges, we can build a bandwidth tree where we can get a maximal fat-tree by gradually removing GPUs that could share upper link bandwidth with other GPUs. On top of Figure~\ref{fig:load-ring}, we get a maximal bandwidth fat-tree (solid red line) from the original bandwidth tree (dashed green line).
Data loading from the host memory happens only on the GPUs that belong to this maximal fat-tree.

\begin{figure}[t]
\centering
 \begin{minipage}[t]{0.93\linewidth}
 \begin{lstlisting}[mathescape]
vertex$^{\ell + 1}$ = CommNet(vertex$^{\ell}$)
  params p = $[W^{\ell}_H,\ W^{\ell}_C]$
  // Passing data over edges
  edge$^{\ell}$ = Scatter(vertex$^{\ell}$);
  // no edge-parallel computation
  acc = ApplyEdge(edge$^{\ell}$)
  $\quad$return edge$^{\ell}$.src
  set Gather.accumulator = sum
  accum = Gather(acc)
  // compute new vertex data
  vertex$^{\ell + 1}$ = ApplyVertex(vertex$^{\ell}$, accum, p)
  $\quad$return $\text{ReLU} \left( \text{p}.W^{\ell}_H \otimes \text{vertex}^{\ell} + \text{p}.W^{\ell}_C\otimes \text{accum} \right)$
  return vertex$^{\ell + 1}$
\end{lstlisting}
\end{minipage}
\vspace{-2ex}
\caption{Communication Neural Net (CommNet)}\label{fig:commnet}%\vspace{-4mm}
\end{figure}

\begin{figure}[t]
\centering
 \begin{minipage}[t]{0.93\linewidth}
 \begin{lstlisting}[mathescape]
vertex$^{\ell + 1}$ = GCN(vertex$^{\ell}$)
  params p = $W^{\ell}$
  // Passing data over edges
  edge$^{\ell}$ = Scatter(vertex$^{\ell}$);
  // edge.data refers to the static weight
  acc = ApplyEdge(edge$^{\ell}$)
  $\quad$return edge$^{\ell}$.src $\times$ edge.data
  set Gather.accumulator = sum
  accum = Gather(acc)
  // compute new vertex data
  vertex$^{\ell + 1}$ = ApplyVertex(vertex$^{\ell}$, accum, p)
  $\quad$return $\text{ReLU} \left( \text{p}.W^{\ell} \otimes \text{accum} \right)$
  return vertex$^{\ell + 1}$
\end{lstlisting}
\end{minipage}
%\vspace{-2ex}
\caption{Graph Convolutional Network (GCN)}\label{fig:gcn}%\vspace{-4mm}
\end{figure}
%\begin{figure}[t]
%\centering
%\begin{minipage}[t]{0.95\linewidth}
%\begin{lstlisting}[mathescape]
%CommNet-VertexProgram(u,$\ell$)
%// scatter edge data
%  scatter$\left(D^{\ell}_{u}\right) $
%  // no edge computation
%  update$\left(D^{\ell}_{u},\ D^{\ell}_{v}, \ e_{uv}, \ W_{uv}^{\ell} \right) $
%  $\quad$return $D^{\ell}_{u}$
%  // gather edge data
%  gather$\left(D^{\ell}_{uv}\right)$
%  sum$\left({m}^{\ell}_{u},\ \hat{m}^{\ell}_{u} \right)$:
%  $\quad$return ${m}^{\ell}_{u}+\hat{m}^{\ell}_{u}$
%  // vertex neural network
%  apply$\left(D^{\ell}_{u},\ m^{\ell}_{u},\ [W^{\ell}_H,\ W^{\ell}_C]\right)$:
%  $\quad$return ReLU$ \left(W^{\ell}_H \otimes D^{\ell}_{u}\ +\ W^{\ell}_C\otimes {a}^{\ell}_{u} \right)$
%\end{lstlisting}
%\end{minipage}
%%	\vspace{-2mm}
%\caption{Communication Neural Net}\label{fig:commnet}%\vspace{-4mm}
%\end{figure}

\begin{figure}[t]
\centering
 \begin{minipage}[t]{0.93\linewidth}
 \begin{lstlisting}[mathescape]
vertex$^{\ell + 1}$ = Max-pooling GCN(vertex$^{\ell}$)
  params p = $[W^{\ell},\ W^{\ell}_{pool},\ b^{\ell}]$
  // Passing data over edges
  edge$^{\ell}$ = Scatter(vertex$^{\ell}$);
  // NN computation using source data
  acc = ApplyEdge(edge$^{\ell}$,p){
  $\quad$return $\sigma\left(\text{p}.W^{\ell}_{pool} \otimes \text{edge}^{\ell}.\text{src} +\text{p}.b^{\ell} \right)$
  set Gather.accumulator = max
  accum = Gather(acc)
  // compute new vertex data
  vertex$^{\ell + 1}$ = ApplyVertex(vertex$^{\ell}$, accum, p)
  $\quad$return $\text{ReLU} \left(\text{p}.W^{\ell} \otimes \text{accum}\right)$
  return vertex$^{\ell + 1}$
\end{lstlisting}
\end{minipage}
%\vspace{-2ex}
\caption{Max-Pooling GCN (MP-GCN)}\label{fig:graphsage}%\vspace{-4mm}
%\vspace{-1ex}
\end{figure}

\begin{figure}[t]
\centering
\begin{minipage}[t]{0.93\linewidth}
\begin{lstlisting}[mathescape]
vertex$^{\ell + 1}$ = GG-NN(vertex$^{\ell}$)
  //different for each edge type
  params p, A
  edge$^{\ell}$ = Scatter(vertex$^{\ell}$);
  // edge.data refers to edge type
  acc = ApplyEdge(edge$^{\ell}$, A){
  $\quad$return $\text{A}(\text{edge}^{\ell}.\text{data})\ \otimes\ \text{edge}^{\ell}.\text{src}$
  set Gather.accumulator = sum
  accum = Gather(acc)
  // compute new vertex data using GRU
  vertex$^{\ell + 1}$=ApplyVertex(vertex$^{\ell}$, accum, p)
  $\quad$return GRU(vertex$^{\ell}$, accum)
 return vertex$^{\ell + 1}$
\end{lstlisting}
\end{minipage}
%\vspace{-2ex}
\caption{Gated Graph Neural Network (GG-NN)}\label{fig:ggnn}%\vspace{-4mm}
%\vspace{-1ex}
\end{figure}

%\begin{figure}[t]
%\centering
%\begin{minipage}[t]{0.9\linewidth}
%\begin{lstlisting}[mathescape]
%vertex$^{\ell + 1}$ = GG-NN(vertex$^{\ell}$){
% params p = $[W_z,\ W_r,\ U_z,\ U_r,\ W,\ U]$
% //different for each edge type
% params A
% edge$^{\ell}$=Scatter(vertex$^{\ell}$);
% // edge.data refers to edge type
% acc = ApplyEdge(edge$^{\ell}$){
% $\quad$return $\text{A}(\text{edge}^{\ell}.data) \otimes \text{edge}^{\ell}.src$
% }
% set Gather.accumulator = sum
% accum = Gather(acc)
% // compute new vertex data using GRU
% vertex$^{\ell + 1}$=ApplyVertex(vertex$^{\ell}$, accum, p){
% $\quad {z}^{\ell+1} = \sigma \left(\text{p}.W_{z}\otimes \text{vertex}^{\ell} \  +\ \text{p}.U_{z}\otimes \text{accum} \right)$
% $\quad {r}^{\ell+1} = \sigma \left(\text{p}.W_{r}\otimes  \text{vertex}^{\ell} \  +\ \text{p}.U_{r}\otimes \text{accum} \right)$
% $\quad h^{\ell+1} = \tanh\left(\text{p}.W\otimes(\text{vertex}^{\ell}\odot{r}^{\ell+1})+\text{p}.U \otimes \text{accum}\right)$
% $\quad$return $(1-{z}^{\ell+1})\odot\text{vertex}^{\ell}\  +\ {z}^{\ell+1}_{u} \odot h^{\ell+1}_{u}$
% }
% return vertex$^{\ell + 1}$
% }
%\end{lstlisting}
%\end{minipage}
%%	\vspace{-2mm}
%\caption{Gated Graph Neural Network (GG-NN)}\label{fig:ggnn}%\vspace{-4mm}
%\end{figure}

\section{Applications}\label{subsec:apps}
\pn{} can support many different types of graph-based neural networks~\cite{ggcn,googlegnn,defferrard2016convolutional,henaff2015deep,hamilton2017inductive,kipf2016semi,li2015gated,encoding,commnet2016}.
In this section, we present the layer-wise programs of several representative GNN models from
the literature.
% which exhibit different patterns of learning.  

%We choose these examples as they represent different computation patterns.

\para{Communication neural net (CommNet)}\cite{commnet2016} is a model where cooperating agents learn to communicate among themselves
before taking actions. Each agent is controlled by a deep feed-forward network, which additionally
receives the summed transmissions of other connected agents. This network can be used
to solve multiple learning
communication tasks like traffic control.
In CommNet, \emph{there is no computation on the edge}, so the \emph{ApplyEdge} stage is simpy a passthrough (see Figure~\ref{fig:commnet}). Each agent gets the summed transmissions via the \emph{Gather} stage, and generates its
new hidden state in the \emph{ApplyVertex} stage.

\para{Graph convolutional networks (GCN)}\cite{henaff2015deep,kipf2016semi} generalizes the notion of CNNs to an operation
that operates on an arbitrary graph. This algorithm has been
applied in many semi-supervised or unsupervised
graph clustering problems, such as entity
classification in a knowledge graph.
In GCN, \emph{there is computation (without neural networks) on the edge for weighted neighbor activation}.
In the GCN program (see Figure~\ref{fig:gcn}), the \emph{Scatter} operation feeds
vertex features into the \textbf{ApplyEdge} function, which multiplies it by the static edge weight determined by the vertex degree.
The \emph{Gather} stage returns the weighted sum of activations of
neighbors on which the \emph{ApplyVertex} stage applies a fully-connected layer.
%a graph convolution layer
%propagates information to vertices from their neighbors by
%computing a weighted sum of
%neighbors' activations. Then, a fully-connected layer is applied
%on all the vertices, with a shared weight matrix.

%\emph{GraphSAGE} is a inductive method that learns node
%representations through the aggregation of neighborhood information. It supports several aggregation function such as
%mean or pooling.
\para{Max-pooling GCN (MP-GCN)}\cite{hamilton2017inductive} applies the max-pooling operator to each of
the computed features, which can effectively capture different aspects of the neighborhood set.
In MP-GCN, \emph{there is an NN-based computation on the edge with the max aggregation instead of mean}. In its program (see Figure~\ref{fig:graphsage}), the \emph{Scatter} stage passes the source
vertex feature vector into fully-connected neural
network defined in \textbf{ApplyEdge} function. Then the \emph{Gather} stage returns the element-wise max of features
of neighbors on which the \emph{ApplyVertex} stage applies a fully-connected layer.
%Each neighbor's vector independently is through a fully-connected neural
%network and an elementwise max-pooling operation is applied to aggregate
%information across the neighbor set.

\para{Gated GCN (G-GCN)}\cite{ggcn,encoding} is our running example which incorporates
the gate mechanism into GCN, so that the model can learn
which edges are more
important for the learning target. Its computation pattern differs from that of Max-pooling GCN
in that \emph{the NN-based computation on edges requires the feature vectors of both source and destination vertices}.
The \emph{Scatter} stage must propagate the feature data of both vertices on the edge (see Figure~\ref{fig:ggcn}).

\para{Gated graph neural networks (GG-NN)}\cite{li2015gated} applies recurrent neural networks (RNNs) for walks on a
graph structure and unroll the recurrence
for a fixed number of steps. This model was used for NLP tasks
and also in quantum chemistry for fast organic molecule
properties estimation. GG-NN has NN-based edge computation, but using different parameters for different edge labels
(the model assumes discrete edge types). Also, GG-NN has dense computation on vertices where the \textbf{ApplyVertex} function is Gated Recurrent Unit (GRU).
In the GG-NN program (see Figure~\ref{fig:ggnn}), different edges can share different parameters in the \textbf{ApplyEdge} function.

%!TEX root = ../ms.tex
\section{Evaluation}
\label{sec:eval}

We implement \pn{} on top of \tf{}~(v1.7) with about
2,900 lines of C++ code and 3,000 lines of Python code. \pn{} extends \tf{} with a front-end to transform SAGA-NN programs into a chunk-granularity dataflow graph, several (fused) scatter/gather operators for efficient graph propagation, and a ring-based streaming scheduling scheme.

In this section, we present the detailed evaluation results and demonstrate the efficiency and scalability of \pn{}, as well as comparing with a state-of-the-art system, TensorFlow.

\para{Experimental setup.} We evaluate \pn{} on a multi-GPU server, which is equipped with dual 2.6~GHz Intel Xeon E5-2690v4 processors (28 cores in total), 512~GB of memory, and 8 NVIDIA Tesla P100 GPU.
%, and a 100Gbps Infiniband (IB) network adapter (Mellanox MT27700) for interconnection.
The installed operating system is Ubuntu 16.04, using libraries CUDA 8.0, and cuDNN 6.

Table~\ref{tab:data-stats} lists the datasets that we used for evaluation,
which are Pubmed citation network \cite{Prithviraj2008}, protein-protein interaction graphs~\cite{KKMMN2016}, BlogCatalog social network~\cite{Tang}, Reddit online discussion forum~\cite{hamilton2017inductive}, and Wikidata~\cite{wiki:xxx}, respectively.
The column \emph{feature} in Table~\ref{tab:data-stats} represents the size of vertex feature vector, and the \emph{label} column means the number of label classes.
%The \emph{reddit\_small} dataset is random sampled from \emph{reddit\_middle}.
%We constructed three post-to-post graphs from Reddit posts made in different granularity of time periods (e.g., week, month and year), which are used to test the system at different data scales.
We test the performance of our system on the task of vertex classification,
e.g., classifying academic papers
into different subjects in the Pubmed citation dataset, which contains sparse bag-of-words feature vectors for each document and a list of citation links between documents.

%Our evaluation uses 5 popular graph-based neural network applications introduced in Section~\ref{subsec:apps}. However, only CommNet, GCN and GG-NN can directly use the Tensorflow operators due to no or simple computation on the edge, in which cases the propagation can be treated as a sparse multiplication.
%%We set the number of layers $\ell = 2$ in each GNN. 
%We use 2-layer GNN in each application, like Figure~\ref{fig:overview_layers}.
%Performance numbers in experiments are average of 10 epochs.
%||||||| .r343
%Our evaluation uses 5 popular graph-based neural network applications introduced in Section~\ref{subsec:apps}. However, only CommNet, GCN and GG-NN can be directly using the Tensorflow operators due to no or simple computation on the edge, in which cases the propagation can be treated as the sparse multiplication.
%We set the number of layers $\ell = 2$ in each GNN. All performance numbers in our experiments are
%calculated by averaging over 100 iterations.
%=======
Our evaluation uses 5 popular GNN applications introduced in Section~\ref{subsec:apps}. Note that only CommNet, GCN and GG-NN can be directly supported using the Tensorflow operators due to no or simple computation on the edge, in which cases the propagation can be treated as a sparse multiplication.
We set the number of layers $\ell = 2$ in each GNN. 
All performance numbers in our experiments are
calculated by averaging over 
10 epochs.
%100 iterations.
%>>>>>>> .r347

\begin{table}[]
	\centering
%	\footnotesize
	\small
	\label{tab:dataset}
	\begin{tabular}{l|r|r|r|r}
		\textbf{Dataset} & \textbf{vertex\#} & \textbf{edge\#} & \textbf{feature} & \textbf{label} \\ \hline
		pubmed         & 19.7K    & 108.4K     & 500     & 3     \\
		protein        & 43.5K    & 205.6K     & 29      & 3     \\
		BlogCatalog    & 10.3K    & 668.0K     & 128     & 39    \\
		reddit\_small  & 46.6K    & 1.4M   & 602     & 41    \\
		reddit\_middle & 233.0K   & 23.2M  & 602     & 41    \\
		reddit\_full   & 2.2M & 571.0M & 300     & 50    \\
		enwiki         & 3.2M & 222.1M & 300     & 12    \\
	\end{tabular}
	\caption{Datasets (K: thousand, M: million). }
	\label{tab:data-stats}
%	\vspace{-2ex}
\end{table}

\subsection{Efficient Graph Propagation}
\pn{}'s efficient Scatter and Gatter kernels play an import role in handling sparsity 
%matrix tensors in 
of graph propagation.

\begin{figure}[t]
	\centering
	\includegraphics[width=\linewidth]{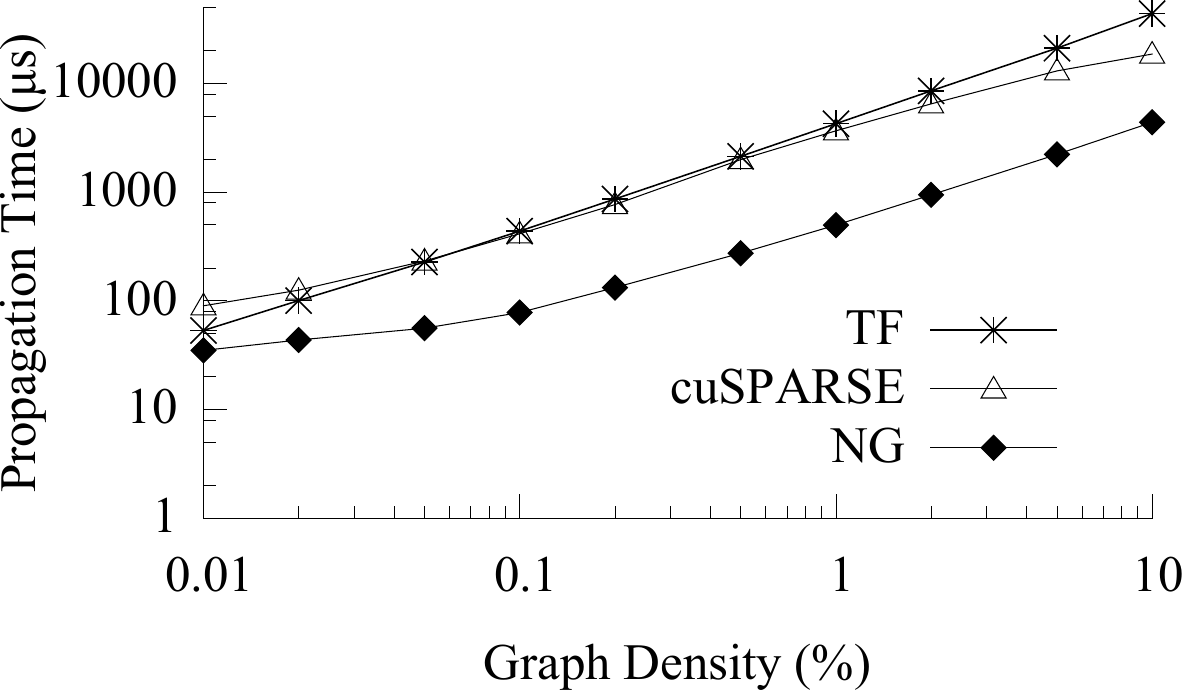}%\vspace{-2mm}
	\caption{Propagation kernel time of TensorFlow(TF), cuSPARSE, and \pn{}(NG) on graphs with different density.}
	\label{fig:kernel_prop}%\vspace{-4mm}
	\vspace{-1ex}
\end{figure}

\para{Micro-benchmark on Synthetic Data.}
To evaluate performance of our propagation kernels, we compare \pn{} with \tf{} and cuSPARSE~\cite{cusparse} on a simple sparse-dense matrix multiplication workload, which can be implemented with SAGA model through just specifying ApplyEdge phase as a multiplication with edge feature.
For \tf{}, we directly use its \texttt{sparse\_tensor\_dense\_matmul} operator.
The inputs of the computation are a $[10,000\times 10,000]$ sparse matrix with variant graph densities (i.e., the percentage of non-zero values) from 0.01\% to 10\%, and a $[10,000\times 128]$ dense matrix.
The performance results are shown in Figure~\ref{fig:kernel_prop}.
Compare to \tf{}, our propagation kernels can speed up by $1.5\times$ to $10\times$. Even compared to cuSPARSE, \pn{} can improve the performance by $1.6\times$ to $6.3\times$. The huge performance gaps are mainly due to  our careful kernel optimization and special GPU threading model design for GNN scenarios, where the vertex data is often a feature vector.
%Compare to \tf{}, our special propagation kernels can speedup the performance by $1.5\times$ to $10\times$. Even compared to cuSparse, NeuGraph can improve the performance by $1.6\times$ to $6.3\times$. The huge performance gaps are mainly because our careful kernel optimization and special GPU threading model design for GNN scenarios, where the vertex data is often a feature vector.
%Compare to \tf{}, our special propagation kernels can speedup the performance by $1.5\times$ to $10\times$. Even compared to cuSparse, NeuGraph can improve the performance by $1.6\times$ to $6.3\times$. The huge performance gaps are mainly due to our careful kernel optimization and special GPU threading model design for GNN scenarios, where the vertex data is often a feature vector.

\para{Real Applications on Small Data.}
To compare with \tf{} on real applications, we need to fit the whole graph data in device memory, as \tf{} cannot support graph larger than single GPU memory. Thus, we just use the first 4 small datasets in Table~\ref{tab:data-stats} to run all the 3 applications that \tf{} supports.
Table~\ref{tab:small} lists the comparison results.
%Overall, \pn{} outperforms \tf{} by ranging from 7.7\% to $2.9\times$ for all applications and datasets.
%Among the 3 applications, the average improvement of GCN, i.e., $1.4\times$, is higher than others. This is mainly because the graph propagation occupies more computation cycles than other applications.
%The improvements are also relative to datasets. For example, the density of \emph{blog} graph is much higher, which leads to higher graph propagation overhead, where \pn{} can speedup. The average improvements of all applications on \emph{blog} dataset is $2.1\times$, while this numbers for the reset datasets are 35.7\%, 23.4\% and $1.1\times$ respectively.
Overall, \pn{} outperforms \tf{} by ranging from 7.7\% to $2.9\times$ among all the applications and datasets.
In the 3 applications, the average improvement of GCN ($1.4\times$), is more than others, mainly because GCN's graph propagation takes more computation cycles than others.
The improvements are also relative to datasets. For example, the density of the \emph{blog} graph is greater, which leads to higher graph propagation overhead, where \pn{} can speed up. The average improvement of all applications on the \emph{blog} dataset is $2.1\times$, while this numbers for the reset datasets are 35.7\%, 23.4\%, and $1.1\times$ respectively.
%Overall, \pn{} outperforms \tf{} by ranging from 7.7\% to $2.9\times$ for all applications and datasets.
%Among the 3 applications, the average improvement of GCN, i.e., $1.4\times$, is higher than others. This is mainly because the graph propagation occupies more computation cycles than other applications.
%The improvements are also relative to datasets. For example, the density of \emph{blog} graph is much high, which leads to high graph propagation overhead, where \pn{} can speedup. The average improvements of all applications on \emph{blog} dataset is $2.1\times$, while the other two are 35.7\% and 23.4\% respectively.
%Overall, \pn{} outperforms \tf{} by ranging from 7.7\% to $2.9\times$ for all applications and datasets.
%Among the 3 applications, the average improvement of GCN, i.e., $1.4\times$, is higher than others. This is mainly because the graph propagation occupies more computation cycles than other applications.
%The improvements are also relative to datasets. For example, the density of \emph{blog} graph is much higher, which leads to higher graph propagation overhead, where \pn{} can speedup. The average improvements of all applications on \emph{blog} dataset is $2.1\times$, while they are 35.7\% and 23.4\% for other two datasets respectively.

\begin{table}[]
	\centering
%	\footnotesize
	\small
	\setlength\tabcolsep{3pt}
	\begin{tabular}{l|r|r|r|r|r|r|r|r}
		\hline
		dataset       & \multicolumn{2}{c|}{pubmed}                       & \multicolumn{2}{c|}{protein}                      & \multicolumn{2}{c|}{blog}                         & \multicolumn{2}{c}{reddit-small}                 \\ \hline
		       & \multicolumn{1}{l|}{NG} & \multicolumn{1}{l|}{TF} & \multicolumn{1}{l|}{NG} & \multicolumn{1}{l|}{TF} & \multicolumn{1}{l|}{NG} & \multicolumn{1}{l|}{TF} & \multicolumn{1}{l|}{NG} & \multicolumn{1}{l}{TF} \\ \hline
		GCN     & 8.2                     & 13.6                    & 14.8                    & 20.7                    & 8.4                     & 32.5                    & 44.2                    & 113.3                   \\ \hline
		CommNet & 14.2                    & 18.6                    & 27.4                    & 33.5                    & 10.6                    & 35.8                    & 62.4                    & 132.4                   \\ \hline
		GG-NN   & 37.6                    & 41.9                    & 77.7                    & 83.7                    & 23.6                    & 49.4                    & 127.3                   & 195.3                   \\ \hline
	\end{tabular}
	\caption{Iteration time (ms) comparison with \tf{}.}
	\label{tab:small}
%	\vspace{-2ex}
\end{table}

\begin{figure}[t]
	\centering
	\includegraphics[width=0.95\linewidth]{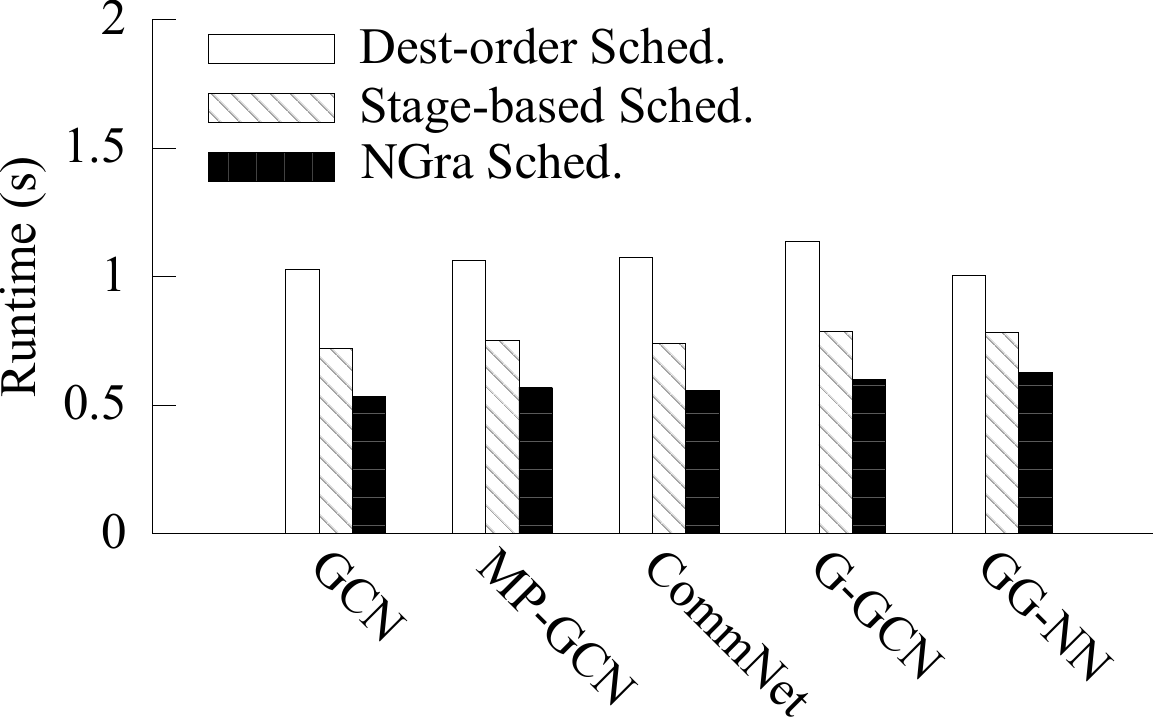}%\vspace{-2mm}
	\vspace{-1ex}
	\caption{Streaming scheduling strategies comparison on different applications. (Data: reddit\_middle) }
	\vspace{-1ex}
	\label{fig:sched}
\end{figure}

\subsection{Scaling-up on a Single GPU}

\begin{figure*}[t]
	\centering
	\includegraphics[width=1\linewidth]{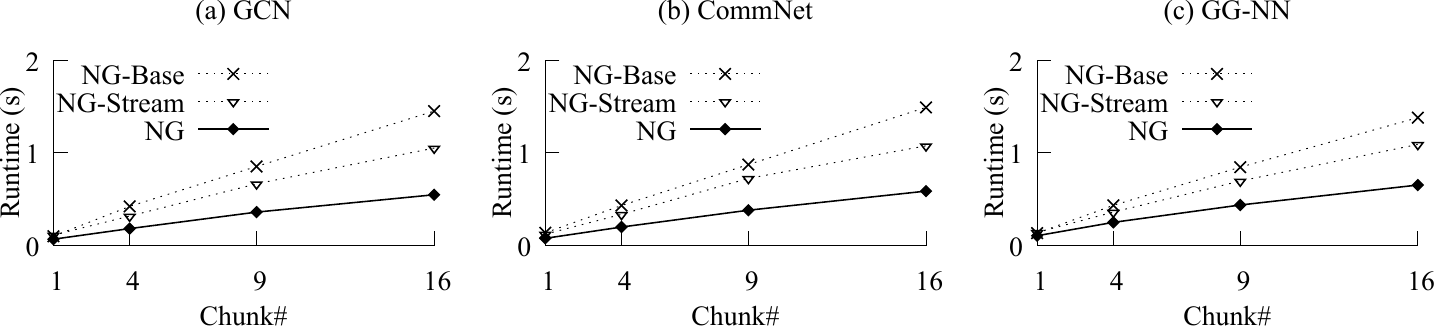}%\vspace{-2mm}
%	\vspace{-1ex}
	\caption{Scaling up performance of \pn{}  on different applications.}
	\label{fig:scale_up}
%	\vspace{3ex}
\end{figure*}

\begin{figure*}[h]
	\centering
	\includegraphics[width=\linewidth]{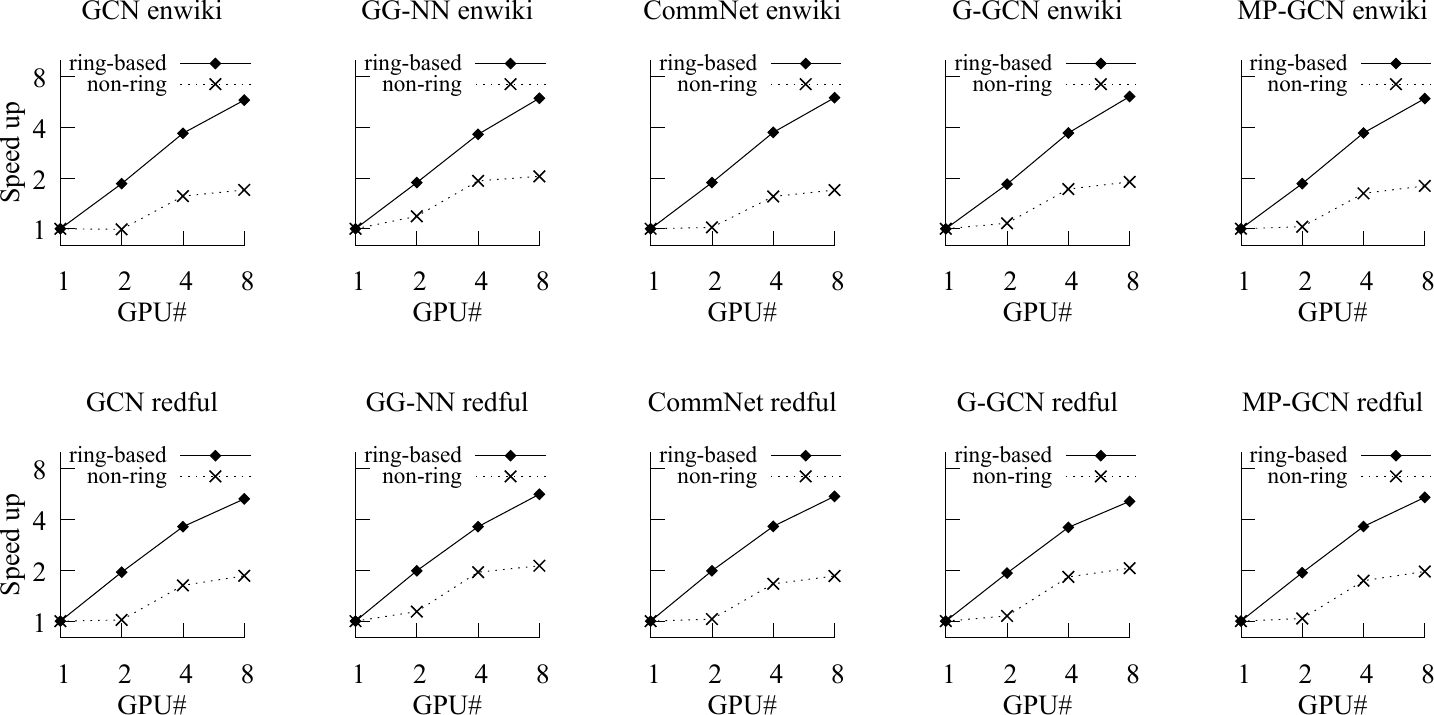}
%	\vspace{-3ex}
	\caption{Speed up of \pn{} with different applications on large graphs.}
	\vspace{-1ex}
	\label{fig:scale_out}%\vspace{-4mm}
\end{figure*}

\pn{} uses the chunk-based streaming mechanism to support graphs that do not fit in GPU memory. We first evaluate different scheduling strategies in this mechanism, as introduced in Section~\ref{sec:system}, and then demonstrate \pn{}'s performance on real applications through comparisons with \pn{}'s baseline versions.

\para{Streaming Scheduling Strategy.}
The scheduling strategy of a chunk-based dataflow graph heavily affects the overall performance, as it determines the number of data swapping introduced in Section~\ref{sec:system:dataflow}.
We demonstrate the benefit of our strategy through comparing with two alternatives: \emph{stage-based} and \emph{dest-order} scheduling strategies. In the \emph{stage-based} strategy, the \emph{Scatter}, \emph{ApplyEdge}, and \emph{Gatter} (S-A-G) are composed as one stage, the \emph{ApplyVertex} as another stage,  where these two stages are executed one-by-one. This will introduce one data swapping between two stages.
In the \emph{dest-order} strategy, 
it prefers to schedule operators in the \emph{Scatter} stage along the direction that destination vertex chunk changes.
%data chunks are enumerated along the order of destination vertices. 
In this case, for each source vertex chunk, the \emph{accum} data needs to be swapped in and out once.
Figure~\ref{fig:sched} shows the comparison results for 5 applications on the \emph{reddit\_middle} dataset.
Compared to the \emph{stage-based} strategy, \pn{}'s scheduling outperforms by 24.9\% to 35.1\% for different applications. For the \emph{dest-order} strategy, \pn{} improves the performance by 60.1\% to 93.1\%. These results demonstrate the benefits of avoiding data-swapping and the importance of \pn{}'s scheduling.

\para{Benefit of Streaming on Real Applications.}
To evaluate the performance gain of streaming in \pn{}, we implement a baseline version of \pn{} by disabling streaming and the optimized graph propagation, denoted as \emph{NG-base}. In this case, \emph{NG-base} can still handle large graph by partitioning it into chunks and processing them sequentially.
We also implement another version through using only chunk-based streaming in \pn{}, denoted as \emph{NG-stream}, which can achieve overlapping of data transmission and computation. We compare end-to-end performance of \emph{NG-base}, \emph{NG-stream}, and \pn{} (\emph{NG}) on 3 applications 
%that \tf{} supports 
in Figure~\ref{fig:scale_up}.
We construct the datasets with different scales by simply duplicating the \emph{reddit\_small} dataset by 1, 4, 9, and 16 copies.
Compared to \emph{NG-base}, \emph{NG-stream} improves the performance by 33.2\%, 29.3\%, and 23.0\%, respectively, for the 3 applications.
%, through just enabling the streaming mechanism.
By further using optimized graph propagation, \pn{} is able to speed up the performance of these applications by $2.5\times$, $2.4\times$, and $1.9\times$ than \emph{NG-base}, respectively.

\subsection{Scaling-out on Multiple GPUs}
\pn{} enables scaling GNN computation to multiple GPUs with ring-based parallel streaming. We compare this mechanism with a baseline without the ring-based strategy, denoted as \emph{non-ring}.
Figure~\ref{fig:scale_out} shows the comparison results for 5 applications on two large datasets, \emph{enwiki} and \emph{reddit\_full}, respectively.
Please note that ring-base mechanism only works on multi-GPU, so 1~GPU data point is the same.
The results show clearly the benefits of ring-based streaming mechanism. For example, when scaling the computation from 1 GPU to 2 GPUs, the average speed up of \emph{non-ring} mechanism is only $1.06\times$, while our \emph{ring-based} one can reach $1.93\times$. This is mainly because, without ring-based design, each of the two GPUs needs to load input data through shared PCIe links concurrently, which easily becomes the bottleneck of the system. The \emph{ring-based} mechanism allows near-linear speed-up because the second GPU can directly load data from the first one, avoiding pressure on the shared upper PCIe links.

From Figure~\ref{fig:scale_out}, we also observe near-linear scalability for our \emph{Ring-based} mechanism before across NUMA nodes.
%, except the case of 8 GPUs. 
%This is due to that current \tf{} do not support NUMA-aware tensor allocation, which leads to suboptimal of \pn{} when reading data cross NUMA node. 
As the current \tf{} implementation can hardly support NUMA-aware tensor allocation well, reading data cross NUMA nodes become suboptimal.
%Nevertheless, our ring-based solution can 
Our further experiments show that we can get $7.09\times$ speed-up on average if we manually enable NUMA-aware tensor allocation.
Generally, \emph{Ring-based} mechanism in \pn{} can improve performance 
%than the \emph{non-ring} one 
by about 2$\times$ on average when using multiple GPUs.

\section{Related Work}
\label{sec:related}

Many real world data can be organized as graph-structured, e.g., web graph, social network, and knowledge graph, etc., where tremendous valuable information can be extracted. A large number of graph processing systems have been proposed and developed to analyze those data through iterative propagation-based computation. Pregel~\cite{Pregel10} first proposes the vertex-program model. It is extended by subsequent work like GraphLab~\cite{GraphLab12} and PowerGraph~\cite{gonzalez2012powergraph} which proposes GAS model to exploit more parallelism on edge-related operations. The GAS model is also well adopted and further extended by a bunch of following work which conduct optimizations on different aspects including graph layout, sequential data access, and secondary storage (e.g., GraphChi~\cite{GraphChi12}, Grace~\cite{Grace12}, XStream~\cite{xstream13}, Chaos~\cite{chaos15}, and FlashGraph~\cite{flashgraph15}), distributed shared memory and RDMA (e.g., Grappa~\cite{Grappa15} and GraM~\cite{Gram15}), NUMA-awareness, scheduling, and load balancing (e.g., Galois~\cite{galois13}, Mizan~\cite{mizan13}, and Polymer~\cite{polymer15}), and graph partitioning (e.g., PowerLyra~\cite{powerlyra15} and BiGraph~\cite{bigraph14}). All these work concentrate on computations conducted on CPU.

There are another series of graph system work that focus on exploiting computation power of GPU for graph processing. Medusa~\cite{medusa} provides simple programming abstractions for GPU-based graph processing and automatically conducts parallel execution on multiple GPUs. CuSha~\cite{cusha} mainly focuses on exploring new graph representations to allow faster graph processing. They both cannot process graphs exceeding the GPU memory capacity. Totem~\cite{totem} statically partitions the graph into GPU and host memory to balance their computation loads, which may not be achievable, especially for large graphs, since the ratios of memory and computation power between GPU and CPU are not aligned. GraphReduce~\cite{graphreduce} can process out-of-memory graphs on a single GPU. It optimizes memory coalescing through using two different formats, the benefit of which can be easily cancelled by the redundant data transfers. GTS~\cite{gts} can also process out-of-memory graphs on multiple GPUs. It does not have mechanisms to avoid redundant vertex data load from host to device memory for multi-GPU case. Garaph~\cite{garaph} exploits edge-centric parallelism and dynamic scheduling to achieve the best performance on the CPU/GPU hybrid platform. Lux~\cite{lux} investigates the placement of graph data over memory hierarchy of CPUs in multiple nodes. Graphie~\cite{graphie} proposes methods to address the challenges when the set of active vertices can change throughout the execution. All these above systems only focus on supporting traditional graph algorithms like PageRank, connected component, and shortest path, etc.

TuX$^2$~\cite{tux2017} pioneers the effort on studying the gap between graph and traditional machine learning computation, while \pn{} moves further to connect graph processing and deep learning which can be well supported by the dataflow frameworks like TensorFlow~\cite{tensorflow16}, PyTorch~\cite{pytorch}, MxNet~\cite{mxnet15}, and CNTK~\cite{cntk14}, etc. A similar past effort is in work of GraphX~\cite{graphx14} which is a graph system built over a general dataflow engine, SPARK~\cite{spark12}, while its target is to connect graph processing with batch-like map-reduce workloads in a single workflow pipeline.

%Most of the existing deep learning frameworks cannot support efficient computations on sparse-structured data through GPU, especially for large graph data. 

\vspace{-2ex}

%!TEX root = ../ms.tex
\section{Conclusion}
\label{sec:con}
GNNs represent an emerging computation model that arises naturally from the need to apply neural network models on large graphs. Supporting efficient and scalable parallel computation for GNN training is demanding due to its inherent complexity. \pn{} is the first to target GNNs, with a new programming abstraction, which is then mapped and optimized as dataflows to execute efficiently on GPUs.

%\section{Graph Execution Engine}
%
%\subsection{}
%
%
%\subsection{Mini-Batch Gradient Descent}
%
%\section{Parallelism}
%
%\subsection{Data Parallel}

{\footnotesize \bibliographystyle{acm}
\bibliography{bibliography}}

%\theendnotes

\end{document}